\documentclass[galaxies,article,accept,pdftex,moreauthors]{Definitions/mdpi}
\usepackage{lineno}

\firstpage{1} 
\pubvolume{11}
\issuenum{6}
\articlenumber{114}
\pubyear{2023}
\copyrightyear{2023}
\externaleditor{Academic Editors: Vladimir Korchagin and Oleg Malkov}
\datereceived{14 July 2023} 
\daterevised{14 October 2023} 
\dateaccepted{31 October 2023} 
\datepublished{16 November 2023} 
\hreflink{https://doi.org/10.3390/\linebreak galaxies11060114} 

\Title{A Study of the Properties and Dynamics of the Disk of Satellites in a Milky-Way-like Galaxy System}

\TitleCitation{A Study of the Properties and Dynamics of the Disk of Satellites in a Milky-Way-like Galaxy System}



\Author{{Xinghai} Zhao $^{1,2,3}$, 
	{Grant J. Mathews} $^{2,3}$\orcidA{},
	Lara Arielle Phillips $^{2,3}$ and
	Guobao Tang $^{2,3,}$*}

\AuthorNames{Xinghai Zhao, Grant J. Mathews, Lara Arielle Phillips and Guobao Tang}

\AuthorCitation{{Zhao}, X.; Mathews, G.J.; Phillips, L.A.; Tang, G.}

\address{%
	$^{1}$ \quad Department of Physics, Dalton State College, Dalton, GA {30720}, USA; xzhao@daltonstate.edu\\
	$^{2}$ \quad Department of Physics and Astronomy, University of Notre Dame, Notre Dame, IN 46556, USA; {gmathews@nd.edu (G.J.M.); lara.a.phillips.127@nd.edu (L.A.P.)}\\
	$^{3}$ \quad Center for Astrophysics, University of Notre Dame, Notre Dame, IN 46556, USA}

\corres{Correspondence: gtang@nd.edu}

\abstract{The dynamics of the satellite systems of Milky-Way-like galaxies offer a useful means by which to study the galaxy formation process in the cosmological context. It has been suggested that the currently observed anisotropic distribution of the satellites in such galaxy systems is inconsistent with the concordance $\Lambda CDM$ cosmology model on the galactic scale if the observed satellites are random samples of the dark matter (DM) sub-halos that are nearly isotropically distributed around the central galaxy. In this study, we present original high-resolution zoom-in studies of central galaxies and satellite systems based upon initial conditions for the DM distribution from the Aquarius simulations but with substantial high-resolution baryon physics added.  We find that the galaxy most like the Milky Way in this study does indeed {contain} a disk of satellites (DOS).  Although one galaxy DOS system does not answer the question of how common such disks are, it does  {allow} the opportunity to explore the properties and dynamics of the DOS system.  Our investigation centers on the spatial arrangement (distances, angles, etc.) of satellites in this Milky-Way-like galaxy system with a specific emphasis on identifying and analyzing the disk-like structure along with its dynamical and morphological properties. 
	Among the conclusions from this study, we find that the satellites and DM sub-halos in the galaxy simulations are anisotropically distributed. The dynamical properties of the satellites, however, indicate that the direction of the angular momentum vector of the whole satellite system is different from the normal direction of the fitted DOS and from the normal direction of the velocity dispersion of the system. Hence, the fitted DOS appears to be comprised of infalling sub-halos and is not a rotationally supported system.}

\keyword{kinematics and dynamics; 
	evolution; 
	dwarf; 
	halos; 
	local group} 

\begin{document}
	
	\section{Introduction}
	
	The alignment of the 11 originally discovered most luminous, or~``classical'', satellites of the Milky Way relative to the Milky Way stellar disk \citep{Lynden-Bell:1976,Kunkel:1976,Kroupa:2005} has been known yet unexplained for decades.  For~example, \citet{Kroupa:2005} fit the spatial distribution of the then available satellite samples into a thin disk that is highly inclined relative to the Milky Way stellar disk. It was then claimed \citep{Kroupa:2005} that this observation is inconsistent with  the hierarchical structure formation  in  $\Lambda CDM$ cosmology on the galactic scale at a high statistical significance level (99.5 percent).   This is true as long as the observed satellites only trace a distribution of the dark matter (DM) sub-halos that reflects the shape of their host DM halo (see \citet{Kroupa:2010} and \citet{Kroupa:2012} for extended reviews). However, this might not be true if, for~example,  the~satellite distribution reflects the asymmetry of filaments in the cosmic~web.
	
	Beyond the apparent alignment of the 11 classical satellites of the Milky Way, similar anisotropic spatial distributions have also been found for the more recently discovered faint satellites of the Milk Way \citep{Metz:2009,Kroupa:2010} (see~\cite{Simon:2019} for a recent review) and the satellites of the Andromeda galaxy (M31) \citep{Grebel:1999,Hartwick:2000,Koch:2006,McConnachie:2006,Metz:2007,Metz:2009,Martin:2009,Richardson:2011,Martin:2013a,Martinez:2022}. 
	
	It has also been suggested that at least some of the satellites have a coherent motion in  phase space \citep{Lynden-Bell:1995,Kroupa:2005,Metz:2007,Metz:2008,Santos-Santos:2020a,Wang:2020,Pawlowski:2021}. However, it is not altogether clear whether this indicates that they may have formed from the same origin or that they arrived~individually.
	
	The spatial distribution and motion of the globular clusters and the streams of stars and gas in the Milky Way have also been analyzed  (e.g.,~\cite{Keller:2012,Pawlowski:2012}).  These results indicate that young halo clusters ($\approx$10--11 Gyr old, and~defined by their horizontal branch morphology and metallicity)
	have a similar spatial distribution to that of the satellites and the streams have similar motion to that of satellites. These findings have further sparked interest in this so-called ``disk of satellites'' (DOS)~\citep{Kroupa:2005,Metz:2007,Metz:2009,Cautun2015,Libeskind2016,Maji2017,Shao2019,Wang:2020,Gu2022} or ``vast polar structure'' (VPOS)~\citep{Pawlowski:2012,Pawlowski:2021} problem that may pose a challenge to  current structure-formation $\Lambda$\emph{CDM} theory~\mbox{\citep{Kroupa:2005,Metz:2007,Kroupa:2010,Kroupa:2012,Pawlowski:2012a,Cautun2017,Pawlowski:2021,Sawala2022}}.
	
	Many efforts have been made to give a natural explanation of the DOS phenomenon. Early studies \citep{Holmberg:1969,Zaritsky:1997} suggested that satellites in disk galaxy systems have a tendency to avoid the equatorial region relative to the disk with a spatial distribution along a certain special direction. {However, data from the 2dF Galaxy Redshift Survey and the Sloan Digital Sky Survey (SDSS) have suggested that this may only be true for host galaxies with low present-day star formation rates (spectral type $\eta < -1.4$)  and for satellites with low radial velocity ($<$160 km s$^{-1}$) relative to the primary}  \citep{Sales:2004,Brainerd:2005,Yang:2005,Azzaro:2007,Bailin:2008,Steffen:2008,Agustsson:2010,Wang:2020}.
	
	Developments in  numerical simulation techniques for galaxy formation and evolution, especially for the baryonic physical processes, have made it possible to study the DOS phenomenon numerically in detail. Early simulations were based on using pure dark matter \citep{Kang:2005} or semi-analytic galaxy formation \citep{Libeskind:2005,Zentner:2005} simulations in a $\Lambda CDM$ cosmology.  Those studies  found that certain satellite systems formed in the simulations are indeed distributed along a special direction that resembles the spatial distribution of the satellites of the Milky Way. They also suggested that the DOS phenomenon may naturally exist in a universe that is described by a $\Lambda CDM$ cosmology~model. 
	
	Subsequently, however, it was argued that, in~contrast to the assumption made by~\cite{Kroupa:2005}, the~distribution of the satellites does not trace the overall distribution of the DM sub-halos. Rather, it is close to that of the sub-halos with the most massive progenitors in the accretion process \citep{Libeskind:2005}. The~direction of the alignment of the satellites has been  found to be close to the  major axis of the host DM halo in the simulations of~\cite{Libeskind:2005,Zentner:2005,Agustsson:2006,Libeskind:2007,Libeskind:2009,Deason:2011,Wang:2012,Wang:2020}. With~a large sample size of 436 and 431 parent haloes, respectively,
	~\citet{Libeskind:2009} and~\citet{Deason:2011} find that statistically the particular type of the spatial alignment of the satellites of the Milky Way is not uncommon (20\%) among similar systems.  However,~\citet{Wang:2012} found that the possibility is much lower ($\sim$1\%).
	
	The possible coherent motion of the satellites in phase space has been investigated by~\cite{Libeskind:2007,Libeskind:2009,Lovell:2011,Deason:2011,Pawlowski:2020,Pawlowski:2021} with simulations. It was found  in their simulations that in some galaxy systems, the direction of the net angular momentum of the satellite system can be close to that of the main galaxy. However, it is rare ($<$10\%) \citep{Libeskind:2009,Deason:2011, Boylan:2021} for satellite systems in the simulations to have an orbiting pole alignment like the one found for the satellites of the Milky Way \citep{Metz:2008}. 
	
	An alternative explanation to the coherent motion of the satellites in the phase space is that the observed satellites are the tidal dwarf galaxies (TDGs) that formed in a close encounter between two galaxies in the Local Group \citep{Zwicky:1956,Lynden-Bell:1976,Kroupa:1997,Kroupa:2005,Metz:2007a,Metz:2008}. The~formation of TDGs in galaxy interactions has been shown to be quite common (e.g.,~\cite{Okazaki:2000,Bournaud:2010}).  Using a large sample of general galaxy interaction simulations, Refs.~\cite{Pawlowski:2011,Pawlowski:2012a} have shown that the resulting TDGs can easily have a similar spatial distribution and orbiting pole to those of the satellites of the Milky Way. Furthermore, using constrained simulations \citep{Wetzstein:2007,Hammer:2010,Yang:2010} of the interaction between the Milky Way and the tidal tail of the Andromeda galaxy,~\citet{Fouquet:2012} found that  the resulting tidal tail matches the spatial distribution and phase space properties of the Milky Way satellites. However, there are still several problems with this theory: First, TDGs are generally believed to contain only baryonic matter (gas) \citep{Barnes:1992,Bournaud:2010}. So, if~a majority of the satellites of the Milky Way are indeed TDGs, they cannot have a very high mass-to-light ratio contrary to  current observations that show a ratio of up to a few thousand (e.g.,~\cite{Simon:2007,Simon:2011}) except in some particular cases \citep{Kroupa:1997,Klessen:1998,Casas:2012}. Second, one needs realistic simulations to study the detailed physical properties of these gas-rich TDGs in the proposed scenarios beyond the ones in~\cite{Pawlowski:2011,Fouquet:2012}. And~finally, the~properties of the proposed galaxy candidates in the galactic interactions simulated in~\cite{Pawlowski:2011,Fouquet:2012} are still in need of close examination as to whether they match the~observations.
	
	On the basis of  previous and new (Gaia DR2) proper motion data of the 11 classical satellites of the Milky Way, Pawlowski and Kroupa~\cite{Pawlowski:2020} showed that there is indeed an alignment of the orbital poles of 7 out of 11 classical satellites. Using the TNG100 simulation data, they argued that such an alignment is extremely rare (0.1\%) in $\Lambda$\emph{CDM} cosmological~simulations.
	
	However, Samuel~et~al.~\cite{Samuel21} used FIRE-2 simulations~\cite{Wetzel23} to show that spatially thin and/or kinematically coherent satellite planes in Milky-Way-like galaxies are {not} very rare at the level of a few percent. They also found that an LMC-like satellite will greatly improve the odds of finding such systems. 
	The conundrum of satellites that are aligned with the DOS but do not appear to be satellites of the LMC has been explored in the context of a single LMC pericentric passage (e.g.,~\cite{Vasiliev:2023a}), and~a second LMC pericentric passage~\cite{Vasiliev:2023b}. The~authors find that the latter scenario leads to a larger debris field and could result in an additional 4--6 dwarf galaxies lost from the Magellanic~system.
	
	A recent  study~\cite{Pawlowski:2021} of  the new proper motion data for M31’s satellite, NGC 147 and NGC 185 has led to the suggestion of the Great Plane of Andromeda (GPoA).  It was shown that these two satellites indeed have an orbital plane aligned with the GPoA. Additionally, they showed that while it is not very rare (more than 4\%) to find two co-orbiting satellites around a Milky-Way-like galaxy in cosmological simulations (e.g., TNG50, TNG100, ELVIS, and~PhatELVIS), it is extremely rare (1 in 1000) for these two co-orbiting satellites to also reside within a thin satellite plane around the main~galaxy.
	
	In another recent work, Sawala~et~al.~\cite{Sawala2022} made use of a collection of zoom-in constrained simulations.  That paper claimed that a thin plane of satellites like that of the Milky Way is relatively common (5.5\%) but short-lived if numerically disrupted satellites are~included.
	
	Also, using semi-analytical galaxy formation models, Gu~et~al.~\cite{Gu2022} found that 4.7\% of galaxy clusters and 13.1\% of MW-mass systems have planes of satellites that are thinner than that of the current Milky Way in their simulations. Although~the satellite accretion is  highly anisotropic at higher redshift, whether it can account for the narrow aspect ratio of the currently observed satellite planes remains~inconclusive.
	
	More recently, using Magneticum Pathfinder simulations and a new plane of satellites finding algorithm, F\"orster~et~al.~\cite{Forster22} found that thin planes of satellites exist in almost all simulated galaxy~systems.
	
	By studying the accretion history of the Local Group in the HESTIA simulation suite, Refs.~\cite{Libeskind20,Dupuy22} also found that satellites in the Local Group accreted along the axis of the slowest collapse of the local~filament.
	
	Most recently, Pham~et~al.~\cite{Pham23} used zoom-in high-resolution dark-matter-only simulations to show that the satellite system of the Milky Way is unusual but still consistent with the $\Lambda$\emph{CDM} model at the {2--3}$\sigma$ level, although~it also depends upon the radial distribution of the~satellites.
	
	Finally, Xu~et~al.~\cite{Xu23} did find in TNG50 simulations the presence of one satellite system around a Milky-Way-like galaxy that closely resembles the Milky Way both spatially and dynamically. Although the majority of satellites seem to accrete along the local filaments, the~satellite plane is transient in~nature.
	
	Here, we similarly examine a DOS system within the context of higher resolution simulations of a Milky-Way-like galaxy system. The~original simulations presented in this paper follow the methodology established by the Aquila Comparison Project \citep{Scannapieco:2009,Scannapieco:2011}.  These adopted the initial conditions from the pure dark matter Aquarius simulations \citep{Springel:2008}. Our investigation focuses on the spatial arrangement of satellites in the Milky-Way-like galaxy system that most resembles the Milky Way, that is the $C$ halo.   Interestingly, in~our simulations, this galaxy contains a DOS system.  This {allow}s the opportunity to conduct a dynamical and morphological analysis in this higher resolution context.  Our specific goal is  to shed light on the nature of the anisotropy, alignment and the potential implications it holds for our understanding of the formation and evolution of the DOS system around a Milky-Way-like galaxy within  the framework of $\Lambda CDM$ cosmology.
	
	This paper is organized as follows: we describe the details of our simulations in Section~\ref{sec2}, the~results are presented in Section~\ref{sec3}, followed by a summary and conclusions in Section~\ref{sec4}.

	\section{The Numerical~Simulations}\label{sec2}

	In order to study the formation and evolution of  Milky-Way-like galaxy disk systems, we utilized original  Smoothed-Particle  hydrodynamics (SPH) simulations that were performed using the Research Computing and Cyber-Infrastrucure unit at Penn State University.  These simulations  {are independent of} but guided by results from  the Aquila Comparison Project \citep{Scannapieco:2009,Scannapieco:2011}.  As in the Aquila project, the~original simulations reported here utilized initial conditions from the pure dark matter Aquarius study \citep{Springel:2008}.  A number of Milky-Way like DM halos with similar mass and evolution history to those of the Milky Way were identified in the Aquarius simulations. In~the work reported here we used the initial condition of the ``C'' halo.  This is because the final main galaxy formed in the Aquila Comparison Project with this initial condition has the best  defined stellar disk \citep{Scannapieco:2009,Scannapieco:2011}.  The simulations reported here utilized sufficient resolution of baryon particles to enable identification of the surrounding dwarf galaxies. Interestingly, in~our analysis, we find that a DOS system exists. Although~the existence of a galaxy with a DOS does not answer the question as to how frequent such systems are, the~existence of this system {allows} an opportunity to study the properties and evolution of the DOS system. That is the purpose of the present~paper.
	
	The simulations  {focused on} a zoom-in high-resolution region surrounding the selected halos. Unlike the Aquarius simulations, which only used dark matter particles, both dark matter and gas particles were used in the SPH simulations. We adopted cosmological parameters from the Aquarius simulations, with~$\Omega_{dm}=0.21$ and $\Omega_b=0.04$ in the high-resolution region. Assuming a $\Lambda CDM$ cosmology, we set the cosmological parameters to $\Omega_m=0.25$, $\Omega_\Lambda =0.75$, $H_0=73$ km s$^{-1}$ Mpc$^{-1}$ and $\sigma_8 = 0.9$. Periodic boundary conditions were applied with a box size of 100 {Mpc} $h^{-1}$ (137 Mpc for $h=0.73$) in co-moving coordinates. The simulations began at $z=127$ with the chosen initial condition {s} and  {were evolved} to the current epoch at $z=0$.
	
	The simulation code used in this study is the Tree-PM SPH code GADGET-3~\mbox{\citep{Springel:2005,Springel:2008}}, which is an improved version of the widely used simulation code GADGET-2 \citep{Springel:2005}. The GADGET-3 code incorporates most of the important baryonic physical processes  {such as} metallicity-dependent cooling, star formation, supernova feedback, black hole physics and a multi-phase gas model. Adopting the notation from the Aquila project, the ``C$-$4'' and ``C$-$5'' refer to the baryonic mass resolution level (cf.~\cite{Marinacci14, Sawala12, Scannapieco:2009}).   As~noted in Table~\ref{tab_sim_par}, in~the zoom-in region of the ``C$-$4'' simulation (higher resolution one),  a~mass resolution of $2.57 \times 10^4 M_\odot$ was achieved for gas and star particles, and~$2.70 \times 10^5 M_\odot$ for dark matter particles. For the ``C$-$5'' simulation ({lower} resolution one),  a~mass resolution of $2.06 \times 10^5 M_\odot$ was achieved for gas and star particles, and~$2.16 \times 10^6 M_\odot$ for dark matter particles.  {Although we present results for the ``C$-$5'' simulations in Figures~\ref{fig_gal_vir} and \ref{fig_gal_disk} as well as Table~\ref{tab_sim_par} below, this is only to demonstrate that the general characteristics do not change when going to higher resolution. However, since our goal is to consider satellites as small as 10$^5 M_\odot$, the ``C$-$5'' resolution would contain fewer than one gas particle, making it inadequate for this study. Hence, we focused mainly on using the ``C$-$4'' simulation for the investigations in the present work. This resolution sufficiently supports a detailed study of low-luminosity satellites. By~sufficient resolution, we mean that with a star particle mass of $\sim$$3 \times 10^4 M_\odot$, halos with $10^6 M_\odot$ contain at least 30 star particles.}
	\begin{figure}[H]
		\includegraphics[trim=0cm 0cm 0cm 0cm, clip=true, angle=0, width=0.3\textwidth]{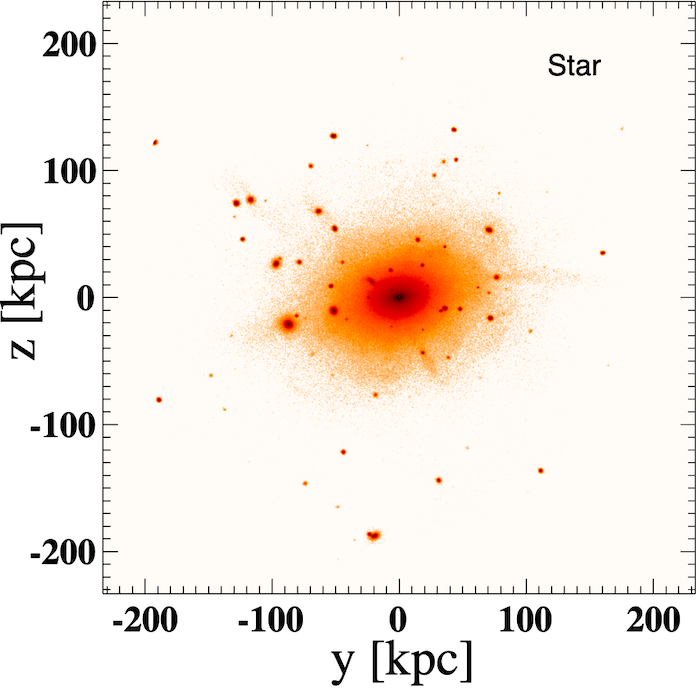}
		\hspace{0cm}
		\includegraphics[trim=0cm 0cm 0cm 0cm, clip=true, angle=0, width=0.3\textwidth]{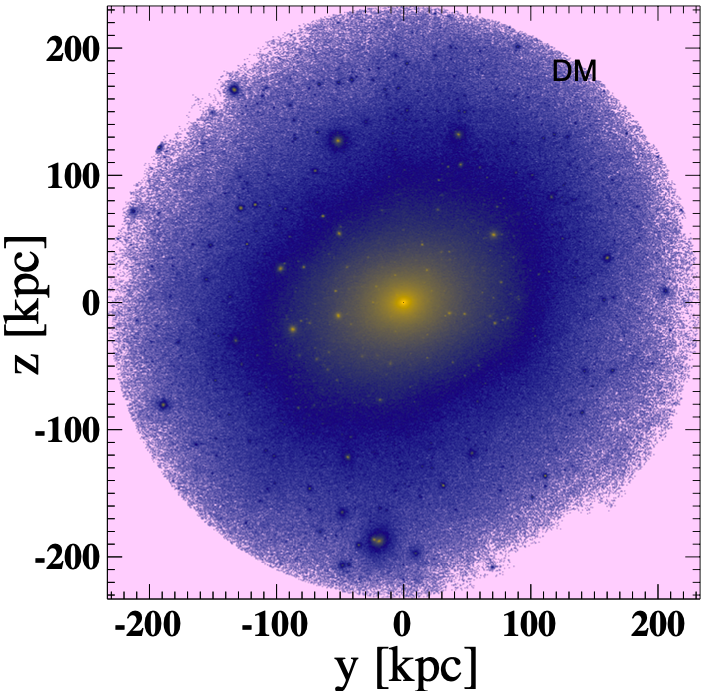}
		\hspace{0cm}
		\includegraphics[trim=0cm 0cm 0cm 0cm, clip=true, angle=0, width=0.3\textwidth]{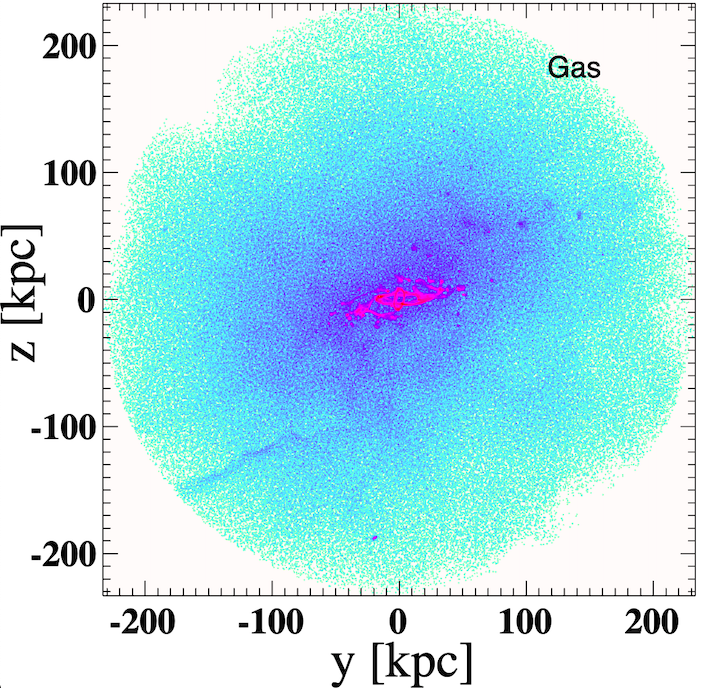}
		\vspace{0cm}
		\caption{{The} edge-on view of the log-scaled density projection map of the stellar, dark matter and gas components in the main galaxy system within its virial radius in the ``C$-$4'' simulation at redshift $z=0$.}
		\label{fig_gal_vir}%
	\end{figure}
	\unskip
	
	\begin{figure}[H]
		\includegraphics[trim=0cm 0cm 0cm 0cm, clip=true, angle=0, width=2in]{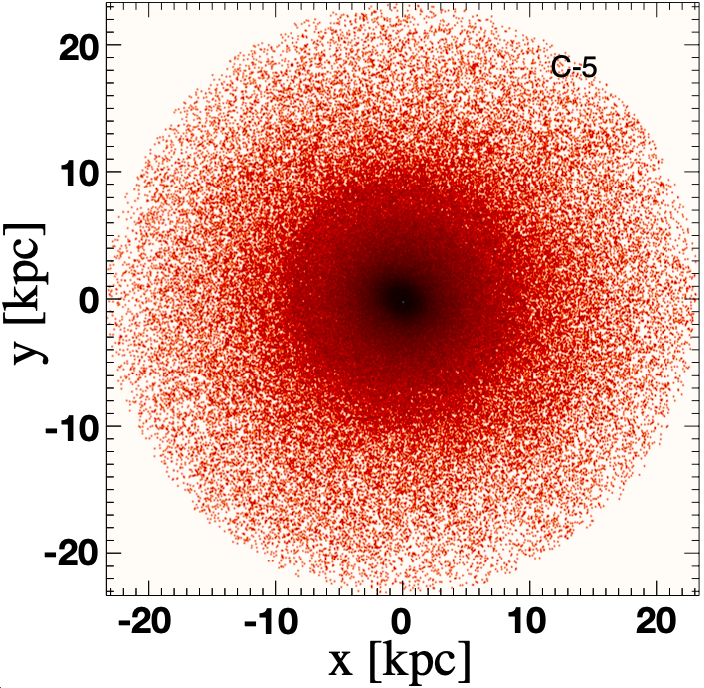}
		\hspace{0.5cm}
		\includegraphics[trim=0cm 0cm 0cm 0cm, clip=true, angle=0, width=1.96in]{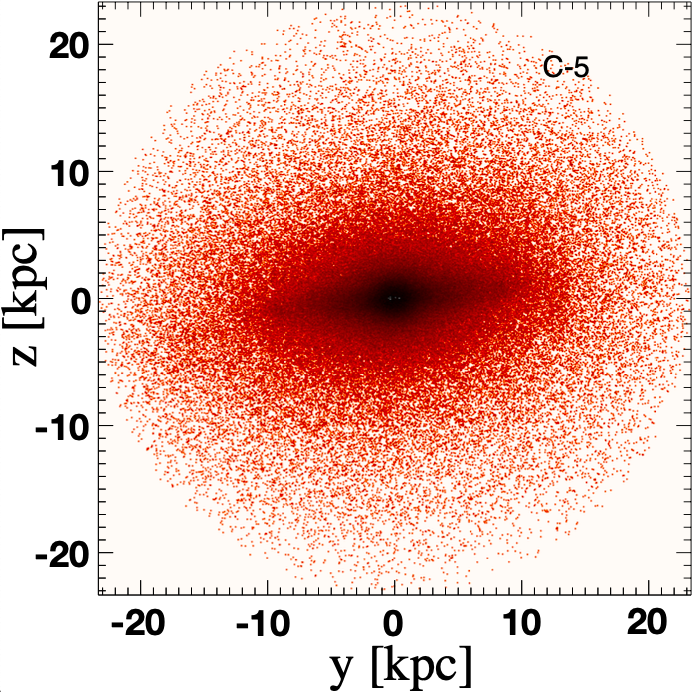}
            \vskip0.25cm
		\includegraphics[trim=0cm 0cm 0cm 0cm, clip=true, angle=0, width=2in]{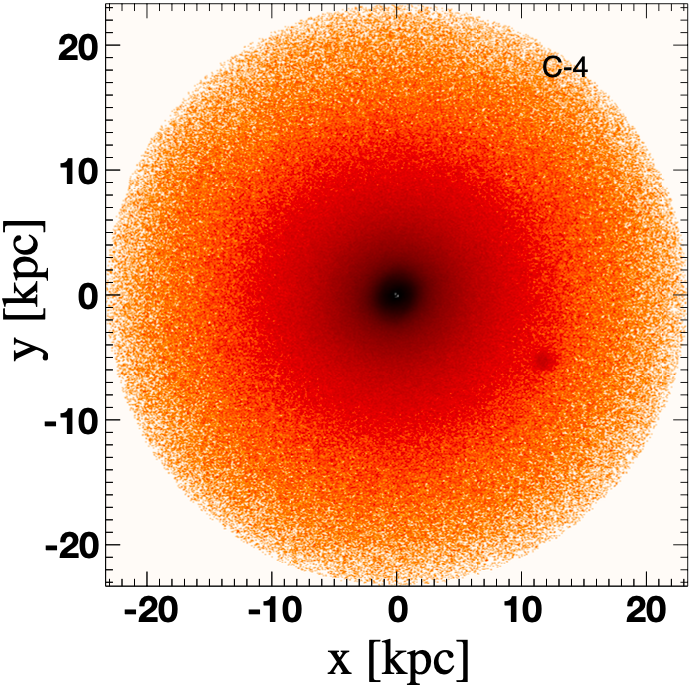}
		\hspace{0.5cm}
		\includegraphics[trim=0cm 0cm 0cm 0cm, clip=true, angle=0, width=2in]{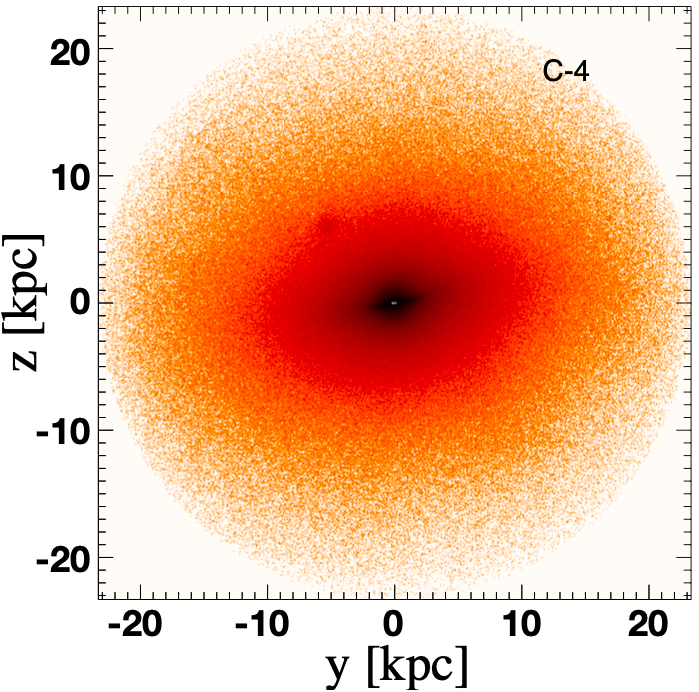}
		\caption{{The} face-on and edge-on views of the density projection map of the stellar component in the central part of the main galaxy in the  ``C$-$5'' and ``C$-$4'' simulations at redshift $z=0$.}
		\label{fig_gal_disk}%
	\end{figure}

	To identify structures at various scales (main halo, sub-halos, central galaxy and satellites) in the simulations, a group finding algorithm was initially applied to the particle data. For~this step, a~friends-of-friends (FOF) algorithm was used to link the dark matter particles based on their mean particle spacing. Subsequently, gas and star particles were linked to the dark matter particle groups using the same algorithm. The~most massive group identified was considered the main galaxy. For~detecting gravitationally bound substructures within the galaxy system, we employed the SUBFIND algorithm \citep{Springel:2001a}. This method first identifies the local over-densities via the SPH kernel interpolation, followed by the iterative removal of gravitationally unbound particles. Any remaining groups with over 20 particles were classified as physically bound substructures. 
	The characteristic parameters of the simulations and the main galaxies at redshift $z=0$ in this study are listed in Table~\ref{tab_sim_par}.


				
    
    \begin{table}[H]
        \caption{List of the characteristic parameters of the simulations and the main galaxies at redshift \( z=0 \).}
        \label{tab_sim_par}
        \normalsize  
        \renewcommand{\arraystretch}{1}
        
  \begin{tabular*}{\hsize}{@{}@{\extracolsep{\fill}}cccccc@{}}
            \toprule
            \multirow{2}{*}{\textbf{Simulation}} &
            \boldmath{$m_{\rm {gas}}$}~\textsuperscript{\textbf{a}} &
            \boldmath{$m_{\rm {DM}}$}~\textsuperscript{\textbf{a}} &
            \boldmath{$m_\star$}~\textsuperscript{\textbf{a}} &
            \boldmath{$\epsilon$}~\textsuperscript{\textbf{b}} &
            \boldmath{$r_{vir}$}~\textsuperscript{\textbf{c}} \\
            & \boldmath{$[10^{4}~{M}_\odot]$} &
            \boldmath{$[10^{5}~{M}_\odot]$} &
            \boldmath{$[10^{4}~{M}_\odot]$} &
            \textbf{\emph{kpc}} &
            \textbf{\emph{kpc}} \\
            \midrule
            C$-$4 & 2.57 & 2.70 & 2.57 & 0.342 & 233 \\
            C$-$5 & 20.6 & 21.6 & 20.6 & 0.685 & 233 \\
\midrule
%
%
            \boldmath{$M_{vir}$}~\textsuperscript{\textbf{d}} &
            \boldmath{$M_{\rm {gas}}$}~\textsuperscript{\textbf{e}} &
            \boldmath{$M_{\rm {DM}}$}~\textsuperscript{\textbf{e}} &            \boldmath{$M_\star$}~\textsuperscript{\textbf{e}} &
            \multirow{2}{*}{\boldmath{$N_{\rm {vir}}$}~\textsuperscript{\textbf{f}}} &
            \multirow{2}{*}{\boldmath{$N_{\rm {1m}}$}~\textsuperscript{\textbf{f}}} \\
            \boldmath{$[10^{12}~{M}_\odot]$} &
            \boldmath{$[10^{10}~{M}_\odot]$} &
            \boldmath{$[10^{12}~{M}_\odot]$} &
            \boldmath{$[10^{10}~{M}_\odot]$} &
            \\
            \midrule
            1.57 & 2.87 & 1.47 & 8.42 & 150 & 263 \\
            1.57 & 4.41 & 1.51 & 7.46 & 50 & 98 \\
            \bottomrule
        \end{tabular*}

		\noindent{\footnotesize{\textsuperscript{a}~{Mass of the gas, dark matter and star particles at high-resolution regions in the simulations.} \textsuperscript{b}~{Gravitational softening length.} \textsuperscript{c}~{Virial radius of the main galaxy defined by the radius of a sphere within which the mean matter density is 200 times the critical density.} \textsuperscript{d}~{Mass enclosed within the sphere of the virial radius.} \textsuperscript{e}~{Mass of the gas, dark matter and stellar components in the main galaxy excluding the satellites.} \textsuperscript{f}~{Number of the luminous satellites around the main galaxy within its virial radius or 1 Mpc radius.}}}
		
	\end{table}

	Figure~\ref{fig_gal_vir} shows the main galaxy formed in the simulations.  Here, we render the edge-on view of the two-dimensional density projection map of the stellar, dark matter and gas components in the main galaxy system within its virial radius in the ``C$-$4'' simulation at redshift $z=0$. In~this figure and throughout the paper, we define the galactic coordinate system of the main galaxy in the simulations by {choosing the direction of the angular momentum of the stellar component within three times the stellar half-mass radius as the normal direction of the stellar disk.} Thus, for~the edge-on view of the galaxy, the~projection is perpendicular to the direction of the stellar component's angular momentum. {It is worth noting that the angular momentum vector is not necessarily perpendicular to the stellar disk, leading to the slightly tilted appearance of the edge-on views in Figure~\ref{fig_gal_vir}.}  In order to show the structures clearly, we only include the particles in the physically bound substructures identified by~SUBFIND. 
	
	As one can see from this figure, by~comparing the distributions of the stellar and DM components in the system, only some of the DM sub-halos have baryonic luminous counterparts, and~thus could be observed as satellites of the main galaxy. Hence, pure dark matter simulations are not sufficient to fully study the DOS phenomenon. The~implementation of baryonic physics is essential for such an investigation. This aspect is related to the ``missing satellite problem'' \citep{Klypin:1999,Moore:1999,Wadepuhl:2011,Font:2011}~(and references therein). However, here we  focus on the spatial distribution{,} not the population of the satellites. Figure~\ref{fig_gal_vir} also shows that the gas component only exists in the main galaxy and that very few satellites appear at redshift $z=0$.  Most satellites are gas~free.
	
	Figure~\ref{fig_gal_disk} shows the face-on and edge-on views of the central part of the main galaxy's stellar component within one-tenth of its virial radius in the ``C$-$5'' and ``C$-$4'' simulations at redshift $z=0$. For~a clear illustration of the structure of the main galaxy, we only include the particles in the main galaxy but not the ones in the satellites. It can be seen from this figure that substantial stellar disk structures are formed in the main galaxies in both of the ``C$-$4'' and ``C$-$5'' simulations. The~galaxy in the ``C$-$5'' simulation has a more distinct disk, similar to the one observed in the previous  ``C$-$5'' simulations \citep{Scannapieco:2009,Scannapieco:2011}, while the galaxy in the ``C$-$4'' simulation exhibits a less distinct disk due to an ongoing minor merger event at $z=0$ in this simulation. {This is  evidenced, for~example, by~the small clump at $(x,y) = (10,-5)$ in the ``C$-$4'' face-on view in Figure~\ref{fig_gal_disk}.  We present both the ``C$-$4'' and ``C$-$5'' simulations here to illustrate  that the fundamental features remain consistent while going to higher resolution. Therefore, for~the remainder of our analyses, we focus on the higher resolution ``C$-$4'' simulations, where the satellite features are more distinctively~defined. }

	\section{Results}\label{sec3}
	
	In this section, we present our results in three parts. First, we examine the spatial distribution of the satellites surrounding the main galaxy. Next, we investigate the dynamical properties of the main galaxy system, analyzing the relative motion of satellites to the main galaxy and the angular momentum across different components in the whole system. For~these analyses, our focus remains on the current system at $z=0$.  {subsequently}, we explore the time evolution of the entire galaxy system to find possible explanations for the findings from the first two~parts.
	
	\subsection{Spatial and Mass~Distributions}
	We first investigate the effects of the satellite luminosity (proxied by mass in the simulation) on the observed anisotropic distribution of the satellites because the early discovered classical satellites of the Milky Way that were included in the previous investigations~\mbox{\citep{Lynden-Bell:1976,Kunkel:1976,Lynden-Bell:1995,Grebel:1999,Hartwick:2000,Kroupa:2005}} (and references therein) are also the more luminous ones. 
	We investigate various satellite samples defined by different lower mass thresholds {($10^8 M_\odot$, $10^7 M_\odot$, $10^6 M_\odot$ and $10^5 M_\odot$)} to determine whether the disk of satellite phenomenon is consistently observed across all samples, indicating its presence as a generic property for the entire satellite population and not solely a result of selection bias from observations. 
	
	{For reference, Figure~\ref{hist} shows a histogram of the  satellite mass function. There is a declining number of satellites with increasing mass, aligning with the expectations of $\Lambda$\emph{CDM} models~\cite{Santos22}. 
		\begin{figure}[H]
			\includegraphics[trim=0cm 0cm 0cm 0cm, clip=true, angle=0, width=3.7in]{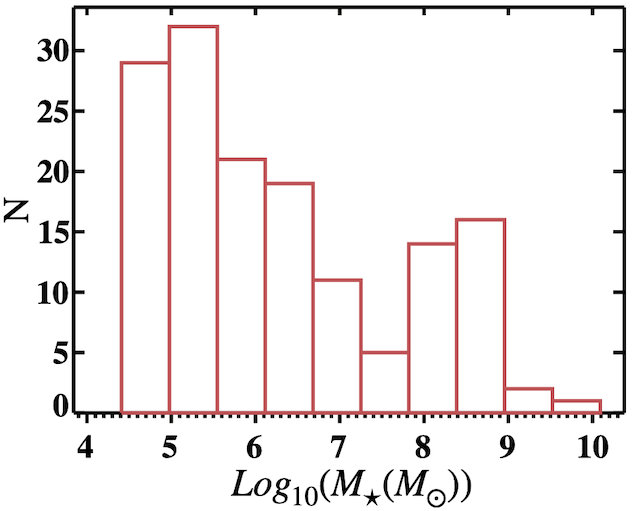}
			\caption{Histogram of the number of satellites in various mass bins from the ``C$-$4''~simulation.}
			\label{hist}%
		\end{figure}
		
		Of course, one should consider the question of  the detectability of satellites over this mass range within the radius of the simulations in this study.  Considering the lowest mass (and therefore lowest luminosity) bin, if~we adopt one solar luminosity for one solar mass, then a 10$^5 M_\odot$ satellite would have an absolute visual magnitude of $-$7.7.  Even for a limiting $G$ band magnitude of 21 as in the GAIA survey~\cite{Gaia:2018} (and ignoring the small $G$-band bolometric correction), these lightest satellites should be detectable out to more than 5 Mpc, which is well beyond the 1 Mpc size of the largest satellite distribution considered below.     }

	In Figure~\ref{fig_dos_vir}, we plot the edge-on views of the main galaxy's stellar disk and the dashed line indicates the orientation of the DOS formed by the satellite samples with  minimum stellar masses of $10^8 M_\odot$, $10^7 M_\odot$, $10^6 M_\odot$ and $10^5 M_\odot$ within the main galaxy's virial radius in the ``C$-$4'' simulation at redshift $z=0$.

	The disk plane of the satellites was fit with the Hesse normal form,
	\begin{equation}
		\vec{n}_{\rm DOS}\cdot\vec{x} - d =0,
	\end{equation}
	where $\vec{n_{\rm DOS}}$ is the unit normal vector of the plane, $\vec{x}$ is the position vector of a point and $d$ is the distance from the plane to the coordinate origin. The fitting uses a least square method on the positions of  {the} satellites relative to the galaxy center as described in~\cite{Kroupa:2005}. With~this method, the~fitted planes in this study are not forced to contain the galaxy center as suggested by~\cite{Metz:2007}. The~galaxy system in Figure~\ref{fig_dos_vir} is oriented in such a way that the  {edge-on view is perpendicular to the angular momentum vector as in Figure~\ref{fig_gal_vir}, but~the projection direction is now rotated so that the fitted DOS plane is always seen edge~on. 
		\begin{figure}[H]
			\includegraphics[trim=0cm 0cm 0cm 0cm, clip=true, angle=0, width=2.2in]{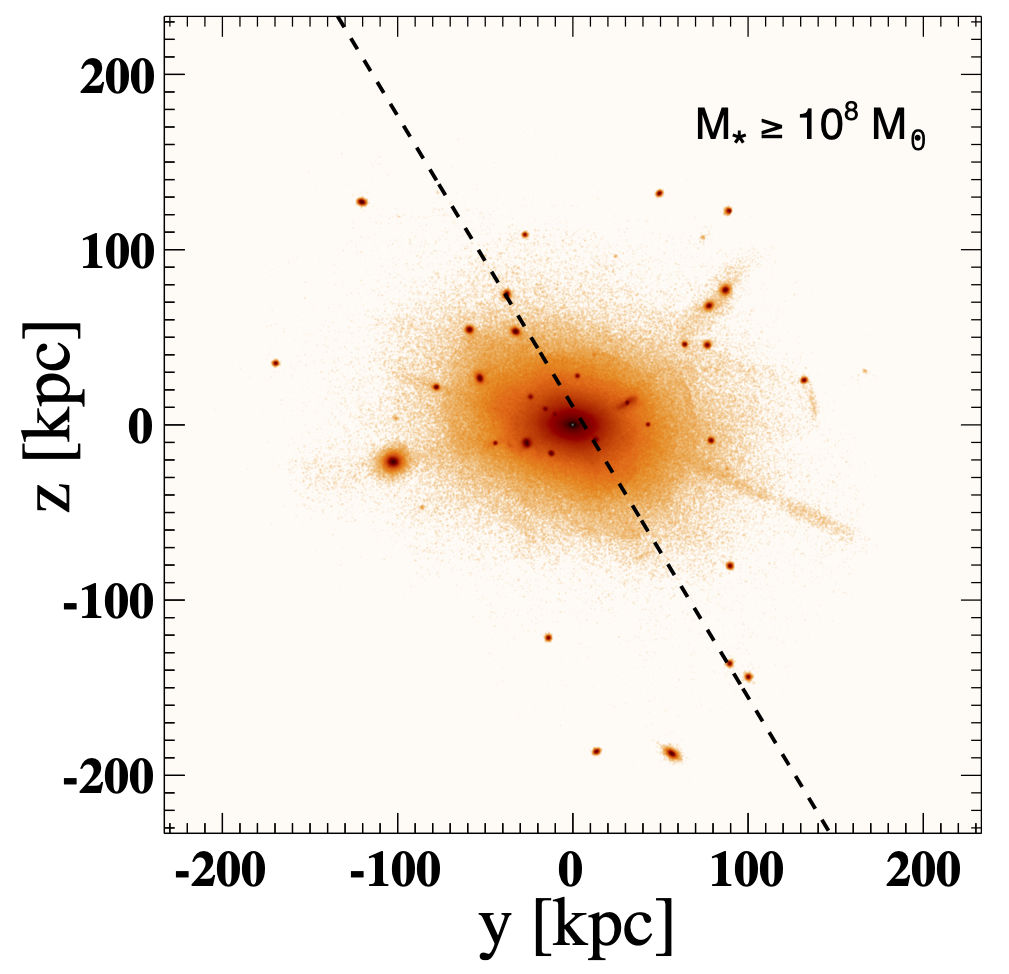}
			\hspace{0cm}
			\includegraphics[trim=0cm 0cm 0cm 0cm, clip=true, angle=0, width=2.2in]{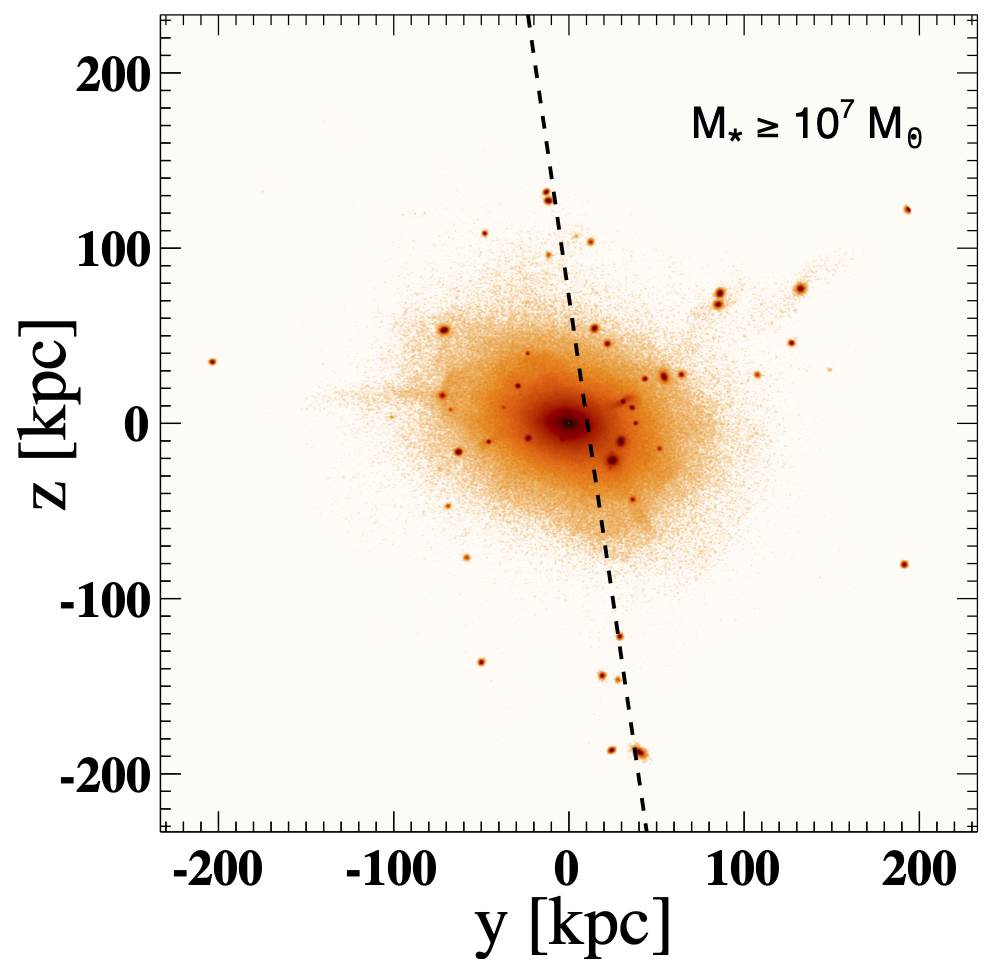}
			\vspace{0cm}\\
			\includegraphics[trim=0cm 0cm 0cm 0cm, clip=true, angle=0, width=2.2in]{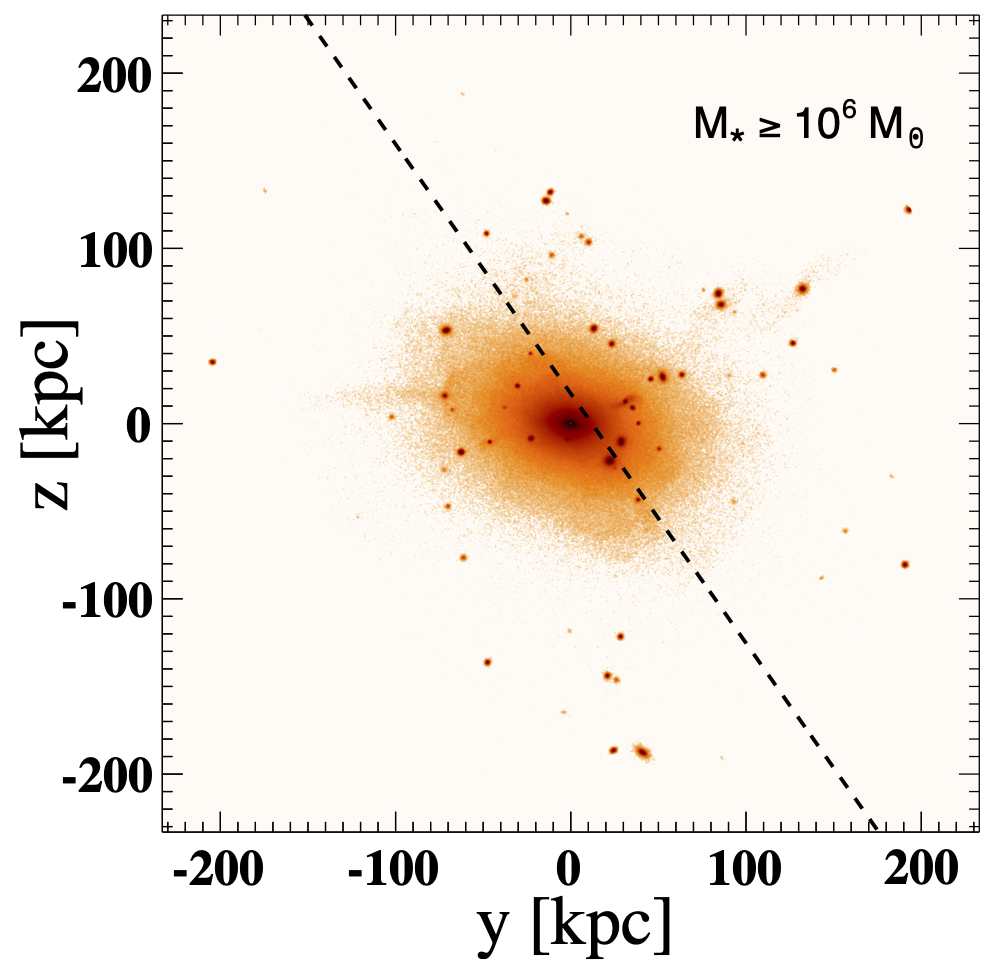}
			\includegraphics[trim=0cm 0cm 0cm 0cm, clip=true, angle=0, width=2.2in]{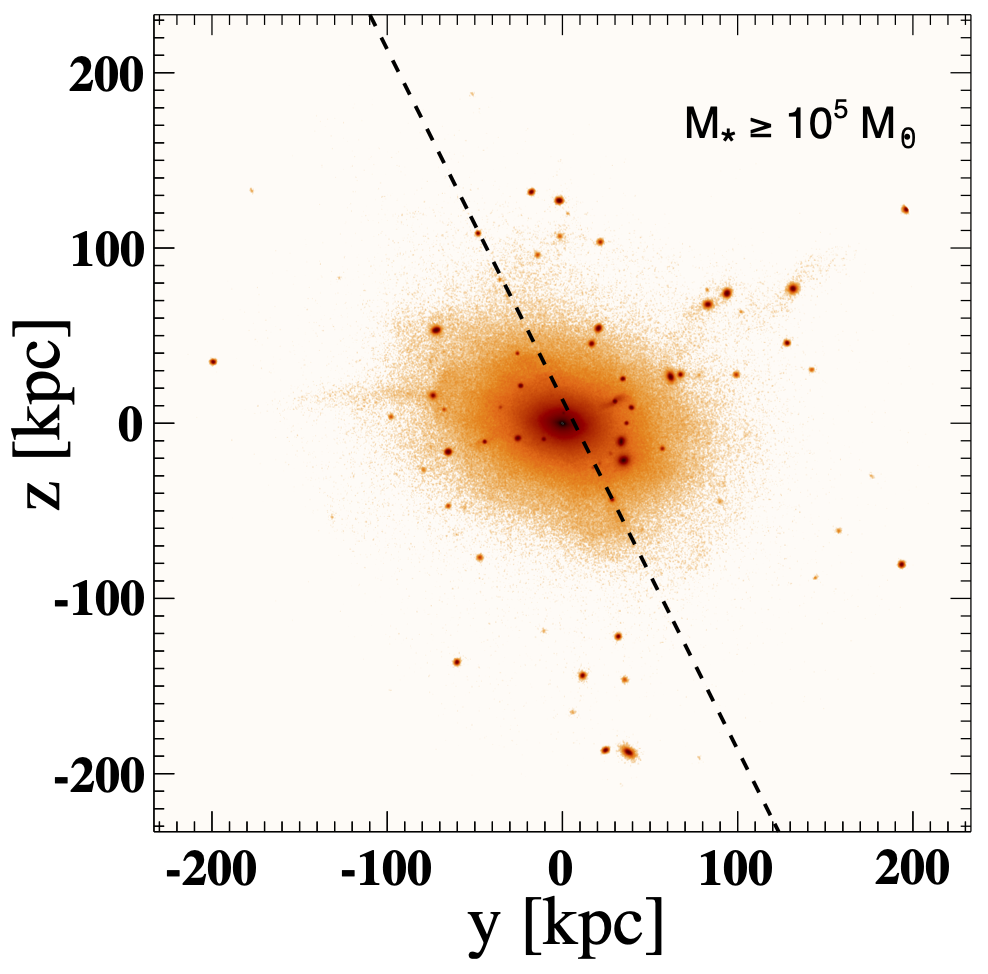}
			\vspace{0cm}
			\caption{{The} edge-on views of the main galaxy's stellar disk within the virial radius in the  ``C$-$4'' simulation at redshift $z=0$. The dashed lines indicate the orientation of the plane of the DOS. The~different DOS data sets are obtained by setting the minimum stellar mass of the satellites to be $10^8 M_\odot$, $10^7 M_\odot$, $10^6 M_\odot$ and $10^5 M_\odot$ as labeled.   }
			\label{fig_dos_vir}%
		\end{figure}

		To offer a parallel perspective in the dark matter environment, in~Figure~\ref{fig_dm_dos_vir}, we also plot the edge-on view of the main galaxy’s  {DM} halo  (defined as perpendicular to the DM halo angular momentum vector) and the DM DOS formed by all  {DM} sub-halos within the virial radius. As in Figure~\ref{fig_dos_vir}, the~dashed line marks the orientation of the DM DOS with the projection direction rotated so that the fitted DM DOS plane is seen edge-on. As~noted below, it is interesting that the orientation of the DM DOS plane is similar to stellar DOS planes shown in Figure~\ref{fig_dos_vir}. }
	
	Figure~\ref{fig_dn_gn_vir} depicts the angle formed between the {normal direction} of the DOS and the {angular momentum direction} of both the main galaxy's stellar disk ({red line}) and the host dark matter halo ({black dotted line}), respectively. This angle is plotted against the minimum stellar mass criterion of the satellite sample. These measurements provide a quantitative assessment of the relative orientation shown in Figure~\ref{fig_dos_vir}. To~determine the orientation of the stellar disk, we first chose the disk stellar particles with similar but simpler criteria than those used in~\cite{Scannapieco:2009}. A~stellar particle is categorized as a disk particle if: (1) it is located inside the main galaxy's disk radius (one-tenth of the main galaxy's virial radius) and outside the galactic {center}\endnote{This is only for the purpose of finding the orientation of the stellar disk in this study, not an exact definition.} (one-tenth of the main galaxy's disk radius); (2) the angle $\theta$ between the angular momentum of the stellar particle and that of the stellar disk satisfies, $\cos\theta \geq 0.7$; and (3) the difference between the stellar particle's circular speed and the circular speed of the disk at the same location is no larger than 30\% of that of the disk. While a more complicated method \citep{Scannapieco:2009} is available, our current  {choice is sufficient}. Our primary goal is to determine the disk's orientation within a few degrees, which is not much affected even if a few particles are~misidentified.
	
	\begin{figure}[H]
		\includegraphics[trim=0cm 0cm 0cm 0cm, clip=true, angle=0, width=3.7in]{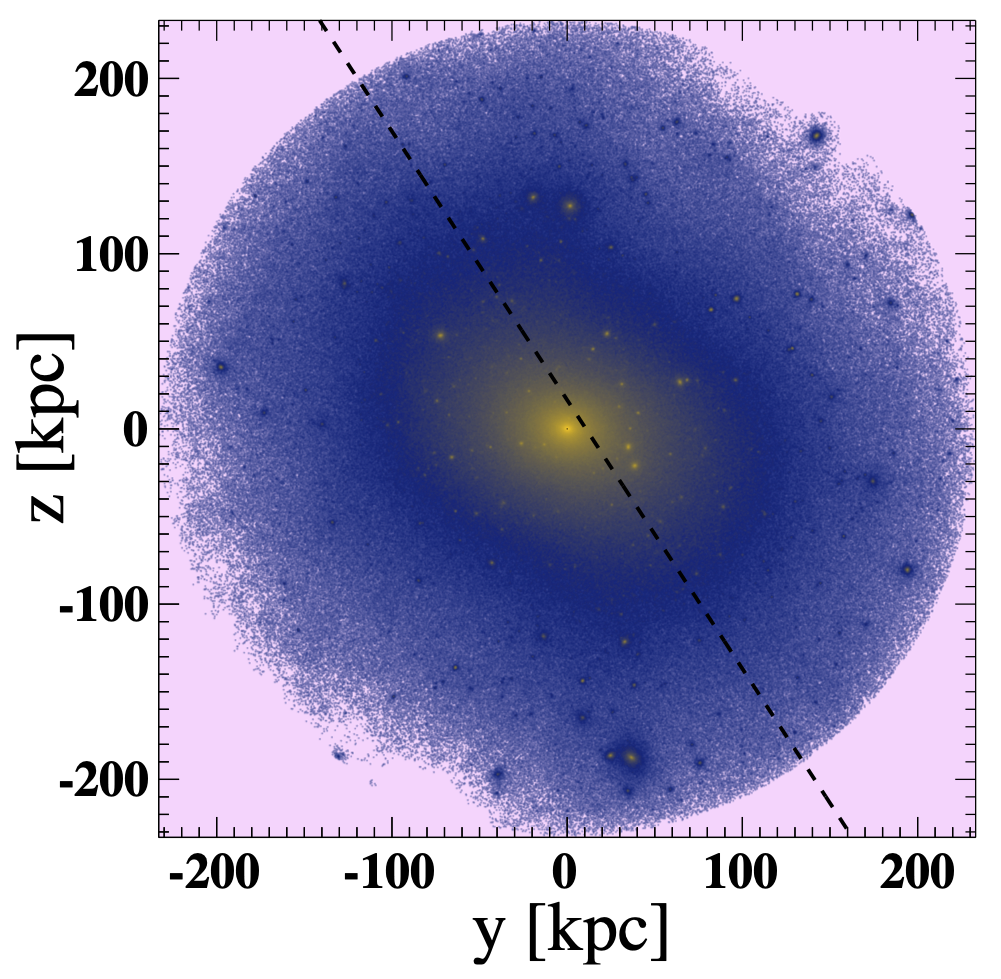}
		\caption{{The} edge-on view of the main galaxy's dark matter halo and the DOS formed by all dark matter sub-halos within the virial radius in the ``C$-$4'' simulation at redshift $z=0$. The dashed line indicates the orientation of the plane of the DOS.}
		\label{fig_dm_dos_vir}%
	\end{figure}
	\unskip
	\begin{figure}[H]
		\includegraphics[angle=0, width=3.7in]{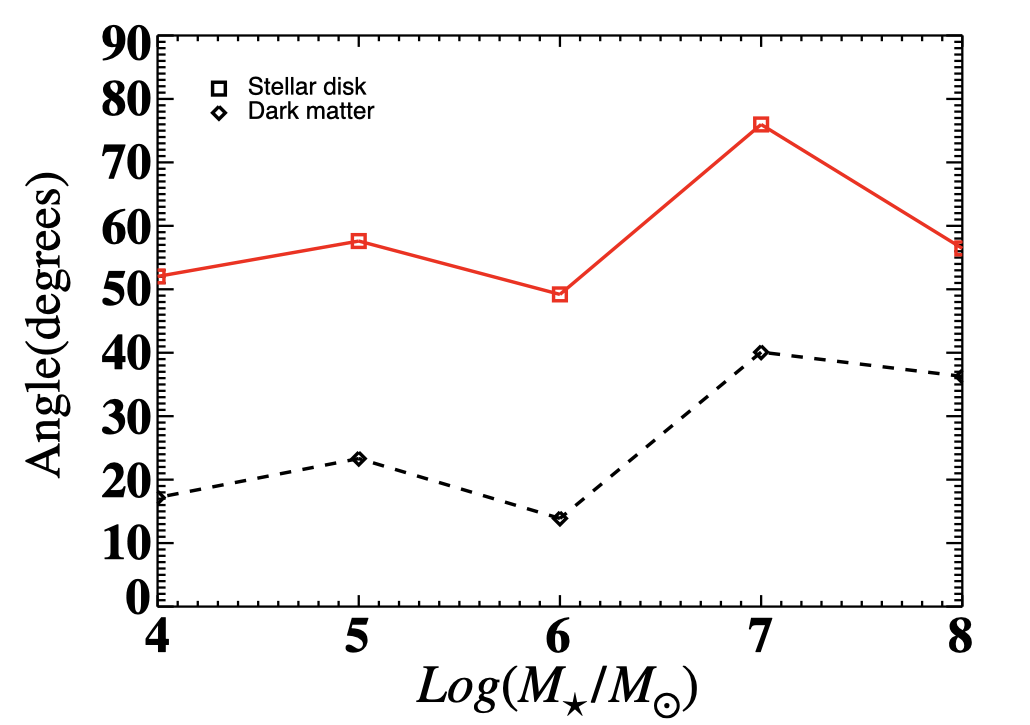}
		\caption{The angle $\theta_{DOS}$ between the normal direction of the DOS  $\vec n_{\rm DOS}$ and the {angular momentum vector $\vec L_{\rm main}$ of }the stellar disk of the main galaxy ({red line}), and~the host dark matter halo ({black dotted line}), respectively. This angle's variation is presented with respect to the minimum stellar mass criterion of the satellite sample within the virial radius in the ``C$-$4'' simulation at redshift $z=0$.}
		\label{fig_dn_gn_vir}
	\end{figure}

	We then diagonalized the second{-}moment tensor of the disk stellar particles,
	\begin{equation}
		I_{ij} = \sum_n x_{i,n}x_{j,n},
	\end{equation}
	to find the orientation of the disk using a similar algorithm to the one that we used to find the orientation of the~DOS.
	
\begin{figure}[H]
		\includegraphics[angle=0, width=4in]{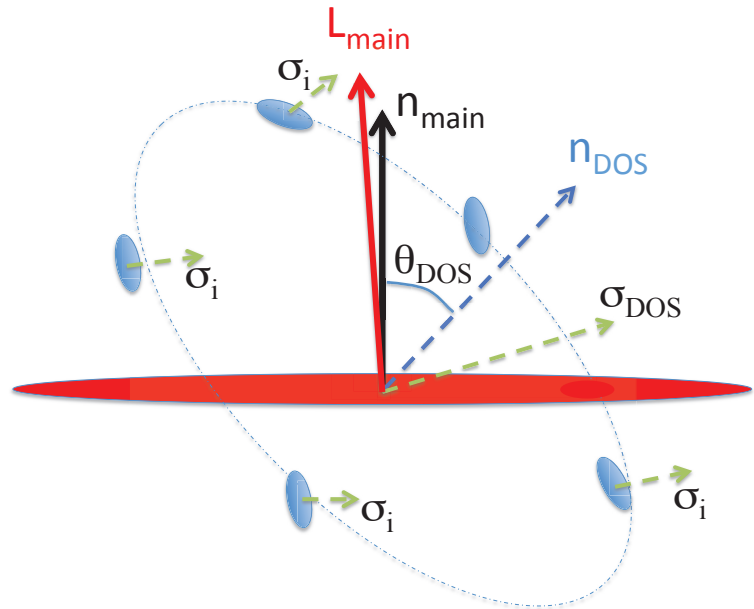}
		\caption{Schematic depiction of the angles  {plotted} in Figures~6, 10, 12 and~14.}
		\label{fig:angles}%
	\end{figure}

	\textls[-5]{Figure~\ref{fig:angles} provides a schematic representation of the various angles depicted in \linebreak \mbox{Figures~6, 10, 12, and 14}}. 
 Here, {we take ${\vec n_{\rm main}} \approx {\vec L_{\rm main}}$ (${\vec L_{\rm main}}$ illustrates the angular momentum vector of the main disk, while ${\vec n_{\rm main}}$ is the direction perpendicular to the stellar disk of the main host)}.  The~quantity $\theta_{\rm DOS}$ approximately represents the angle between ${\vec n_{\rm main}}$ and the normal direction to the DOS, ${\vec n_{\rm DOS}}$.   The~quantity  $\vec \sigma_{\rm DOS} = \sum_i {\vec \sigma_i}$ is the sum of the dispersion vectors (minimum bulk motion) ${\vec \sigma_i}$ of the satellite galaxies in the sample. This direction is determined by summing the velocities of the particles in the satellite.  The~direction maximum sum indicates a bulk flow.  Conversely, the~direction of the minimum sum is only showing the random dispersion and is perpendicular to the direction of any  bulk flow.  Hence, one can use the sum of the dispersion to identify a bulk flow of the~DOS.

	\vspace{-3pt}
	As one can see from Figure~\ref{fig_dos_vir} and quantitatively from Figure~\ref{fig_dn_gn_vir}, the~orientations  of the DOS formed by the satellite samples with different luminosity thresholds are close to each other. In~the luminosity range studied, the~DOSs always have inclination angles\endnote{The inclination angle discussed here is the angle between the DOS plane and the stellar disk plane of the main galaxy. It is equal to the angle between the {normal direction of the DOS and the angular momentum vector direction of the main stellar disk} as plotted in Figure~\ref{fig_dn_gn_vir}.} above about $50 ^{\circ}$ relative to the stellar disk of the main galaxy. For~some particular luminosities (e.g., $M_* > 10^7 M_\odot$), it can be nearly perpendicular to $\vec n_{main}$.
	The angles $\theta_{\rm DOS}$ between the normal direction of the DOS and the normal to the host dark matter halo are much smaller in the same luminosity range than the same angles for the stellar component. This suggests that the DOS exhibits a stronger alignment with the main DM halo compared to the stellar disk of the main galaxy. This alignment seems to imply a tendency for the DOS to be preferentially accreted along the major axis of the DM halo. We speculate that this may point toward the accretion of satellites along filaments. However, this would require a study out to much larger distances in the cosmic~web.

	For the DM sub-halos, we use the same fitting algorithm to find a DM DOS around the main galaxy. {In this case all DM sub-halos with at least 20 particles (i.e., $M_{DM} \sim 10^6~M_\odot$) are counted.  As~shown in Figure~\ref{fig_dm_dos_vir}, this DM DOS has a similar orientation to those of the luminous satellites in Figure~\ref{fig_dos_vir}.  In particular, for the most direct comparison (the plot with $M_* > 10^6 M_\odot$), the~orientations of the DOS indicated by the dashed lines in \mbox{Figures~\ref{fig_dos_vir} and \ref{fig_dm_dos_vir}} are almost identical.} Indeed,  {t}his suggests that the orientation of the DOS formed by the luminous satellites may be close to the preferred direction in the spatial distribution of the dark matter around the main galaxy. These two interesting phenomena
	prompted us to further study the distribution of the satellites and dark matter sub-halos on a larger scale simply because the DM extends to larger~scales.

\begin{figure}[H]
		\includegraphics[trim=0cm 0cm 0cm 0cm, clip=true, angle=0, width=2.2in]{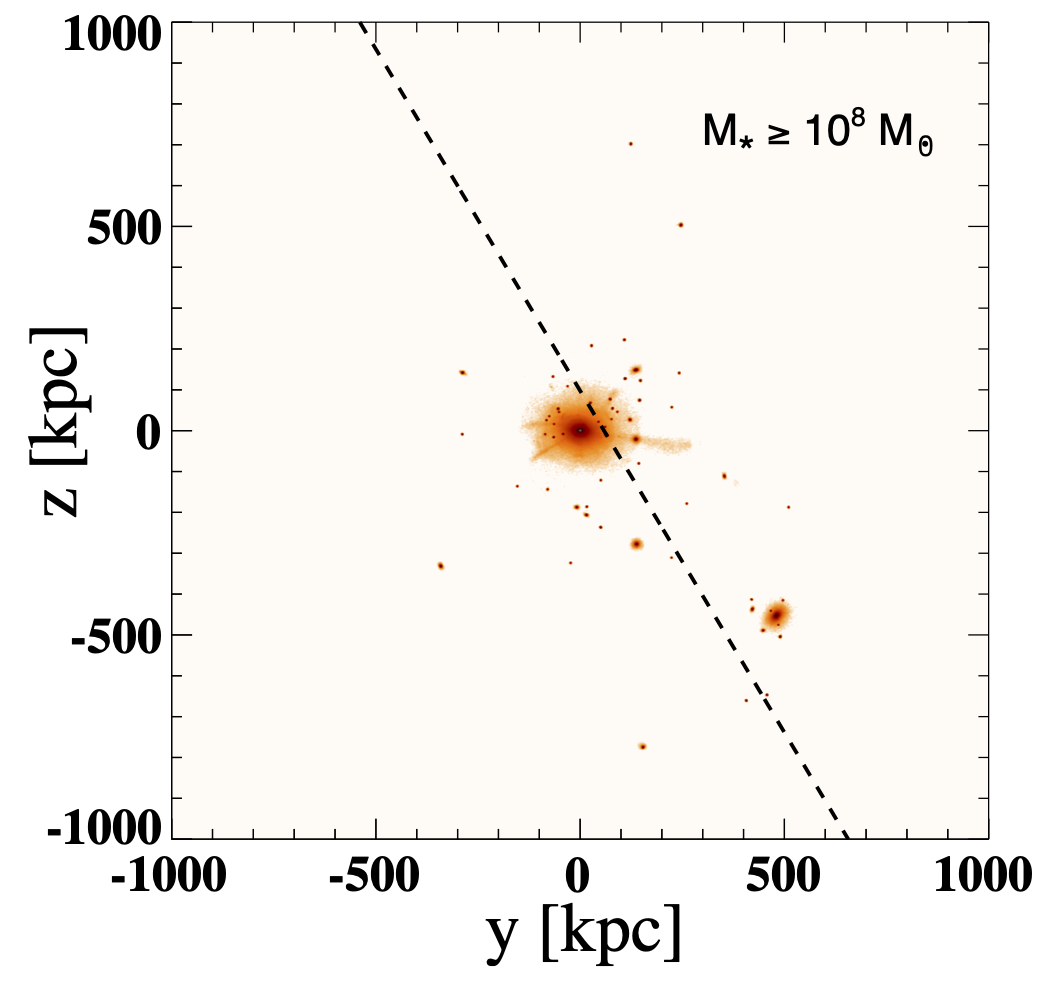}
		\hspace{0cm}
		\includegraphics[trim=0cm 0cm 0cm 0cm, clip=true, angle=0, width=2.15in]{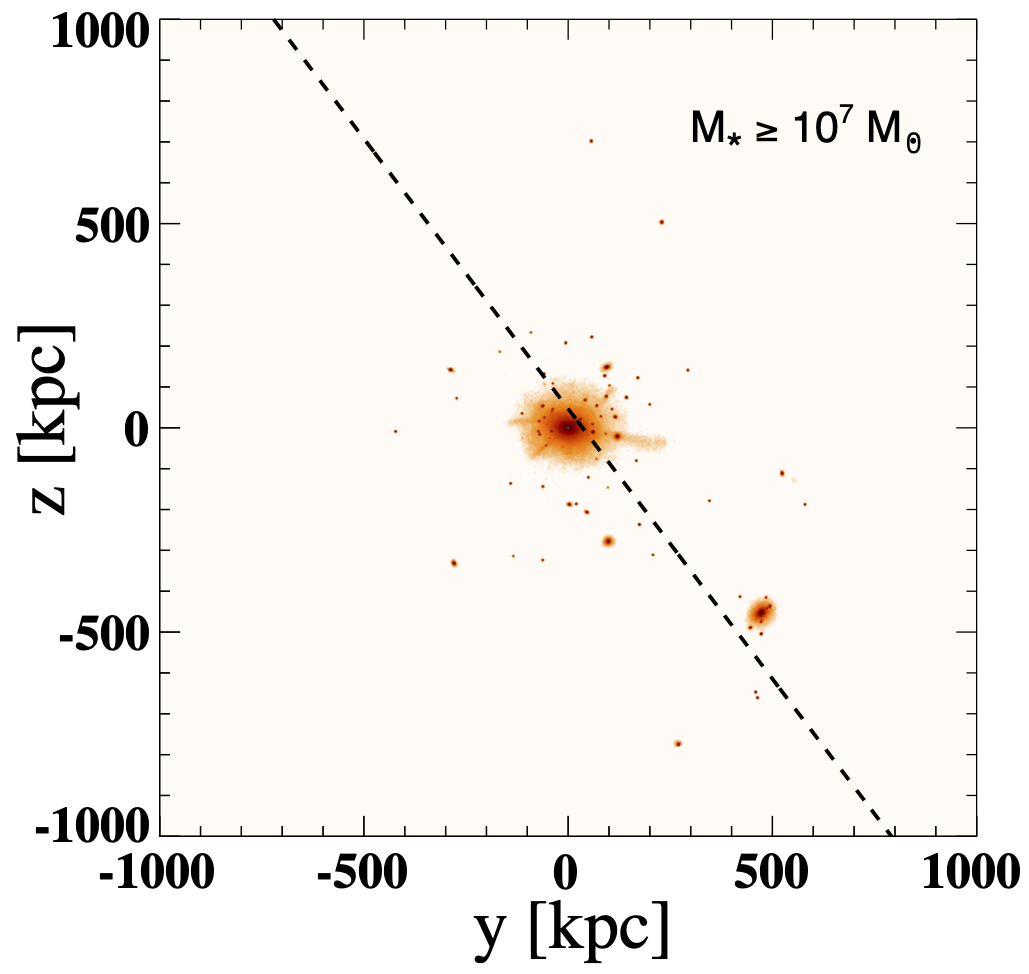} \\
		\vspace{0cm}
		\includegraphics[trim=0cm 0cm 0cm 0cm, clip=true, angle=0, width=2.2in]{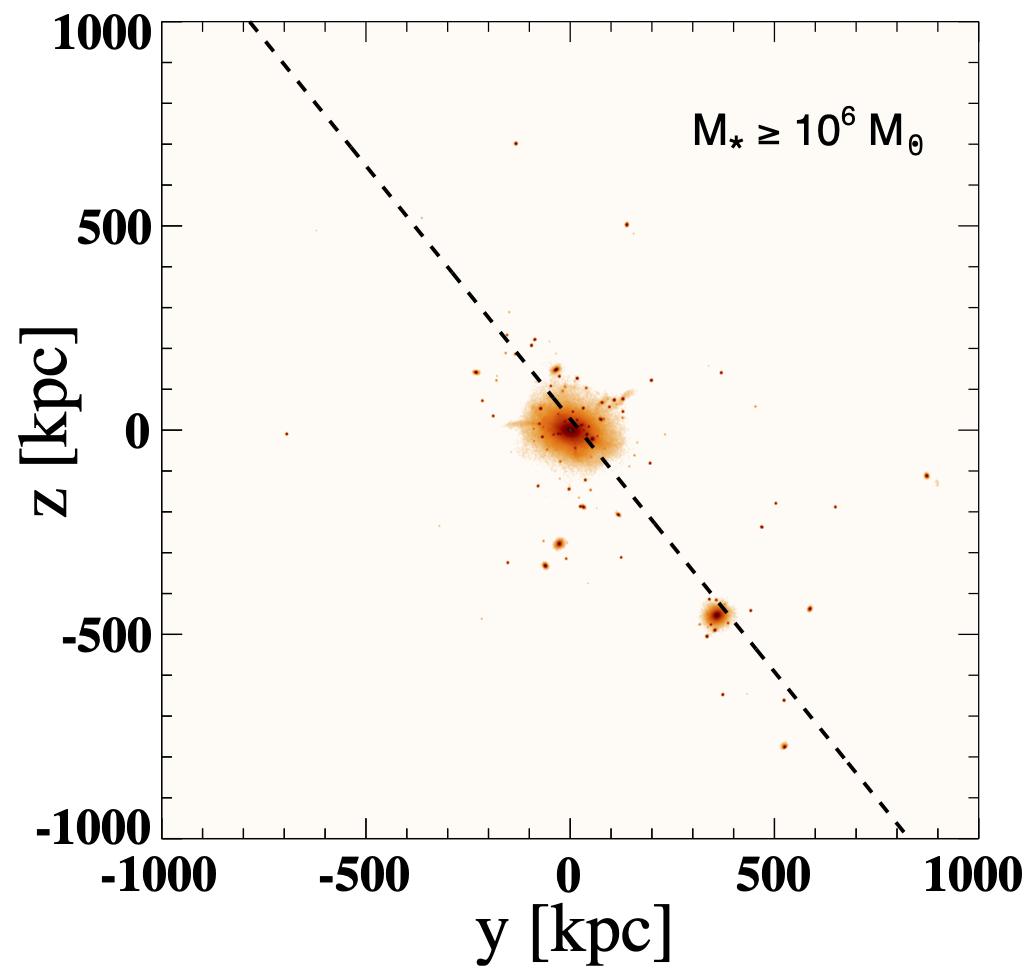}
		\hspace{0cm}
		\includegraphics[trim=0cm 0cm 0cm 0cm, clip=true, angle=0, width=2.2in]{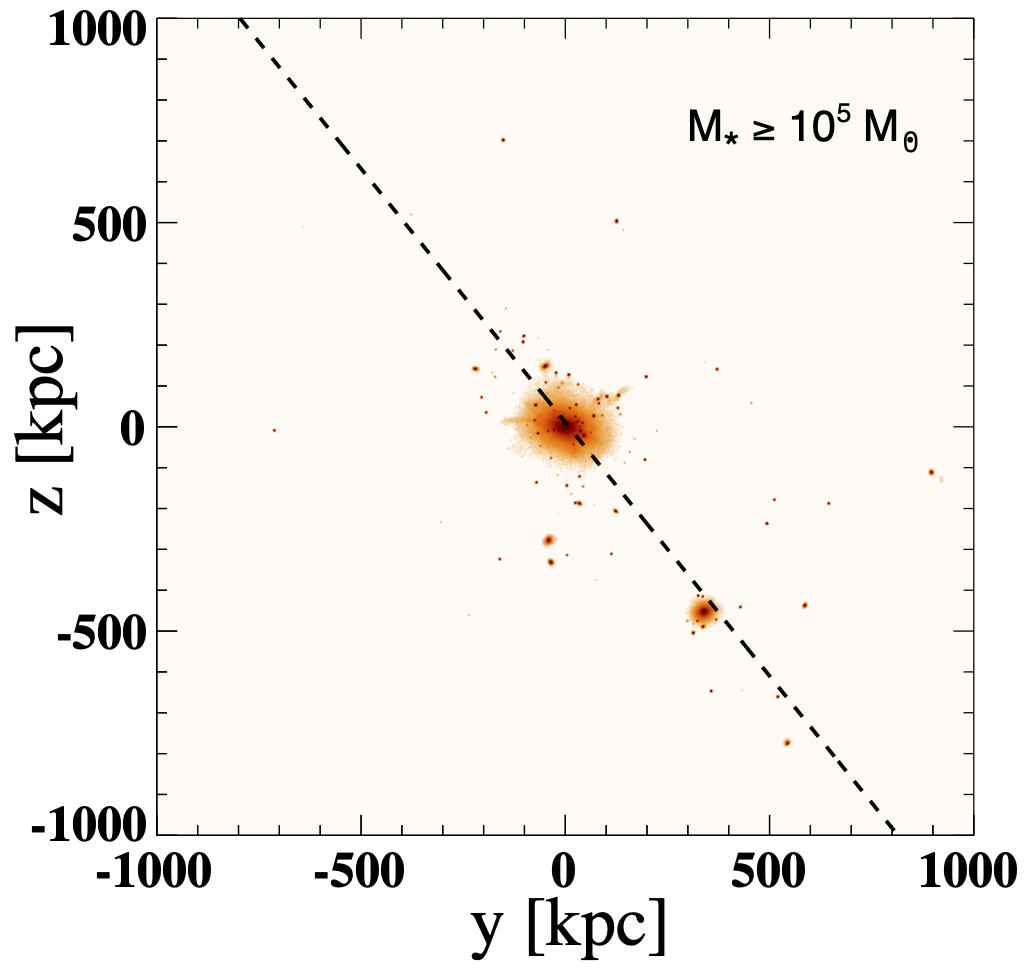}
		\vspace{0cm}
		\caption{{The} edge-on views of the main galaxy's stellar disk and the DOS formed by the satellites with minimum stellar mass of $10^8 M_\odot$, $10^7 M_\odot$, $10^6 M_\odot$ and $10^5 M_\odot$ within 1 Mpc radius in the  ``C$-$4'' simulation at redshift $z=0$. The dashed line indicates the orientation of the plane of the~DOS.}
		\label{fig_dos_1mp}
	\end{figure}

\begin{figure}[H]
		\includegraphics[trim=0cm 0cm 0cm 0cm, clip=true, angle=0, width=3.5in]{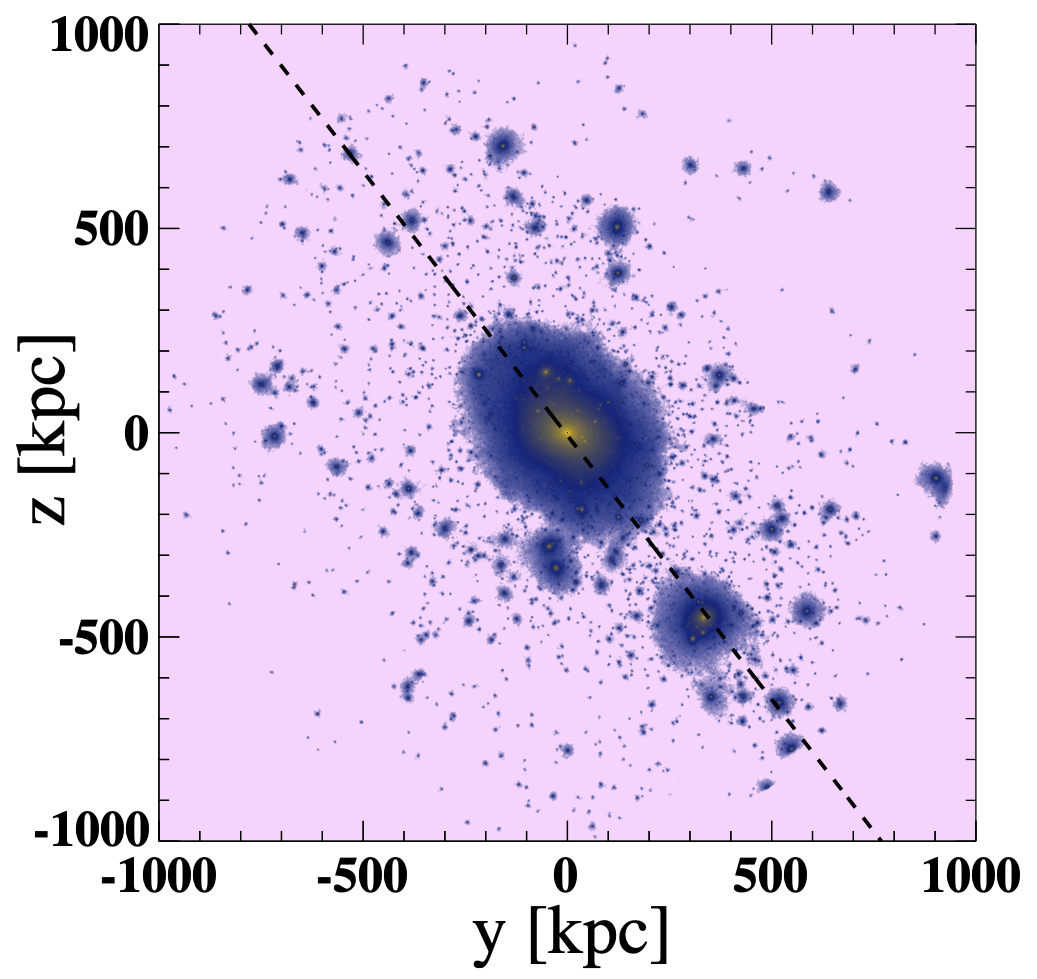}
		\caption{{The} edge-on view of the main galaxy's dark matter halo and the DOS formed by all dark matter sub-halos within 1 Mpc radius in the ``C$-$4'' simulation at redshift $z=0$. The~dashed line indicates the orientation of the plane of the~DOS.}
		\label{fig_dm_dos_1mp}
	\end{figure}

\begin{figure}[H]
		\includegraphics[angle=0, width=4in]{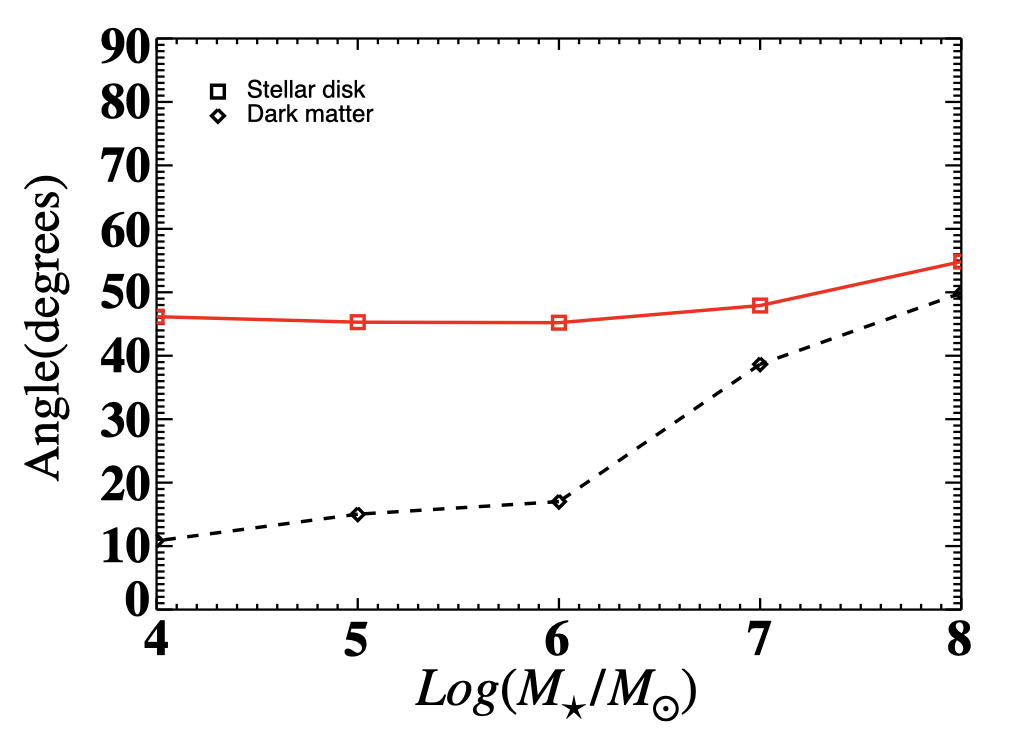}
		\caption{The angle between the normal direction ${\vec n}_{\rm DOS}$ of the DOS and {the angular momentum vector  ${\vec L}_{\rm main}$ of} the stellar disk of the main galaxy  ({red line}) and to the host dark matter halo ({black dotted line}) as a function of minimum stellar mass in the satellites within 1 Mpc radius in the ``C$-$4'' simulation at redshift $z=0$.}
		\label{fig_dn_gn_1mp}
	\end{figure}

	We next investigated the spatial distribution of the satellites\endnote{The satellites studied here may not all be gravitationally bound to the main galaxy, we keep the term only for the purpose of studying their spatial distribution.} within {a} 1 Mpc radius around the main galaxy. We chose this scale because it is comparable to the scale of the Local Group. It is important to acknowledge that{,} as we extend our analysis to a larger radius, the~issue of how distance affects the detectability of satellites becomes more pronounced for observational data. {As noted above, the~satellites discussed here should all be detectable by current surveys out to much more than  one Mpc.} Figures~\ref{fig_dos_1mp}--\ref{fig_dn_gn_1mp} are the identical plots on a 1 Mpc scale to that of the virial-radius scale of Figures~\ref{fig_dos_vir}--\ref{fig_dn_gn_vir}, respectively. As~on the smaller scale of the main galaxy's virial radius, it is clearly shown in \mbox{Figures~\ref{fig_dos_1mp} and \ref{fig_dn_gn_1mp}} that the orientations of the DOS formed by the satellites with different luminosity thresholds are close to each other on a 1 Mpc scale. The~orientations of the DOS formed by the luminous satellites and the DM sub-halos are also very close to each other on a 1 Mpc scale as indicated in Figures~\ref{fig_dos_1mp} and \ref{fig_dm_dos_1mp}. 
	
	This is further illustrated in Figure~\ref{fig_dn_gn_1mp} by the trend of the angle $\theta_{\rm DOS}$ between the normal direction of the DOS and {the angular momentum vector of} the host dark-matter halo. The~angle decreases monotonically from about $50 ^{\circ}$ for $M_\star=10^8 M_\odot$ to about $10 ^{\circ}$ for $M_\star=10^4 M_\odot$. This result suggests that if we fit a complete sample of the satellites with the lowest possible luminosity using the same fitting algorithm, the~fitted DOS plane should be very close to the long axis of the host dark matter~halo.

	\unskip
	
	\unskip

\begin{figure}[H]
		\includegraphics[trim=0cm 0cm 0cm 0cm, clip=true, angle=0, width=4.9in]{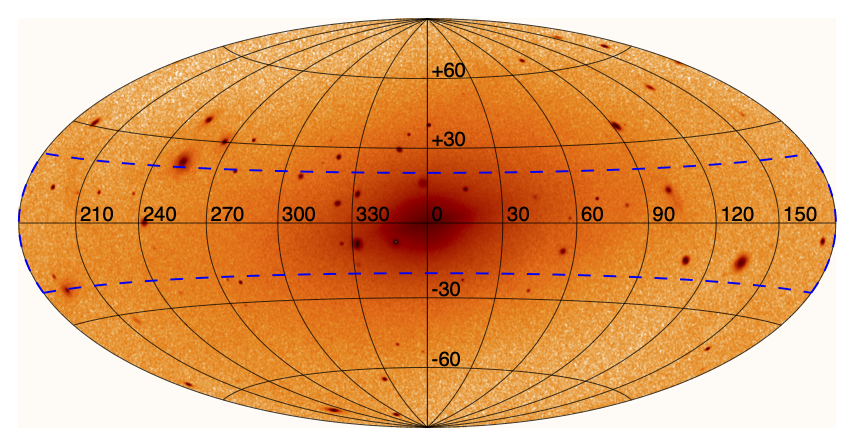} \\
		\vspace{0cm}
		\includegraphics[trim=0cm 0cm 0cm 0cm, clip=true, angle=0, width=4.9in]{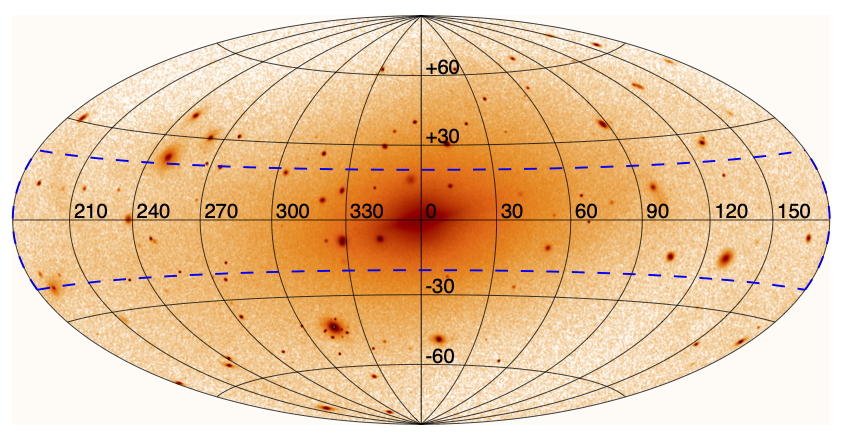}
		\vspace{0cm}
		\caption{{The} Hammer--Aitoff projection of the full sky map of the main galaxy's stellar disk and the satellites with a minimum stellar mass of $10^5 M_\odot$ within the virial radius (\textbf{\boldmath{top}}) and 1 Mpc radius (\textbf{\boldmath{bottom}}) (in RA(deg) and Dec(deg)) in the ``C$-$4'' simulation at redshift $z=0$. The~satellites in the ``Zone of Avoidance'' of the main galaxy are indicated by the dashed~contour.}
		\label{fig_dos_pj}
	\end{figure}

	To compare the spatial distribution of the satellites in our simulations to observations, we plot the Hammer--Aitoff projection of the mock full sky map in Figure~\ref{fig_dos_pj}.  This figure shows the main galaxy's stellar disk and the satellites within the virial radius ({top} plot) and those within 1 Mpc radius ({bottom} plot) in galactic coordinates using our simulation data. For~this sky projection map, the~galactic coordinates are defined by positioning the Sun on a line that passes through the galactic center. This line runs parallel to where the DOS plane intersects the main galaxy's stellar disk. We assume the Sun has a distance of 8.0 kpc from the galactic~center. 
	
	One easily recognizable feature of Figure~\ref{fig_dos_pj} is the almost identical locations of the common satellite points between the virial radius map and the 1 Mpc radius map. This suggests that the DOS formed by the satellites on these two scales are very close to each other. Using the fact that the satellites with low galactic latitudes are often blocked by the Milky Way's own stellar disk and that the observed satellites of the Milky Way ~\cite{Kroupa:2010}  {in} all have galactic latitudes higher than $20 ^{\circ}$. We define the ``Zone of Avoidance'' of the main galaxy in this study as a region with galactic latitudes lower than $20 ^{\circ}$ as marked by dashed lines on Figure~\ref{fig_dos_pj}.
	
	For a quantitative comparison, we list some of the key characteristic parameters of the DOS fitted in our study and those from some observational investigations in Table~\ref{tab_dos_par}{. Their definitions are} explained in the footnote. For~the galactic latitude $b$ of the normal direction of the fitted DOS plane, we notice that the ``C-4 Vir ZOA'' data set (satellites within the virial radius but excluding those within the zone of avoidance) has a $b$ value that is comparable to the values from the fitted DOS of the Milky Way and the ``C-4 Vir'' data set has a $b$ value that is close to the one from the Andromeda galaxy (for further details on the specific characteristics of these data sets, please see the footnote  {below} Table~\ref{tab_dos_par}).

	\begin{table}[H]
		\caption{{List} of the characteristic parameters of the data sets and the fitted DOS at redshift $z=0$. }
		\label{tab_dos_par}
		
		\begin{tabularx}{\textwidth}{cCCCCCCC}
			\toprule
			\multirow{2}{*}{\textbf{Data Set~\textsuperscript{a}}} &
			\boldmath{$R_{\rm {cut}}$}~\textbf{\textsuperscript{b}} &
			\multirow{2}{*}{\boldmath{$N_{\rm {sat}}$}~\textbf{\textsuperscript{c}}} &
			\boldmath{$l$}~\textbf{\textsuperscript{d}} &
			\boldmath{$b$}~\textbf{\textsuperscript{e}} &
			\boldmath{$\Delta$}~\textbf{\textsuperscript{f}} &
			\multirow{2}{*}{\boldmath{$\Delta/R_{\rm {cut}}$}~\textbf{\textsuperscript{g}}} &
			\boldmath{$D_{\rm p}$}~\textbf{\textsuperscript{h}} \\
			
			&
			\textbf{[\emph{kpc}]} &
			&
			\textbf{[deg]} &
			\textbf{[deg]} &
			\textbf{[\emph{kpc}]} &
			&
			\textbf{[\emph{kpc}]} \\
			
			\midrule
			C-4 Vir & 231.6 & 121 & -- & $-$26.6 & 68.6 & 0.30 & 6.0 \\
			C-4 Vir ZOA & 230.8 & 70 & -- & $-$16.0 & 63.5 & 0.28 & 9.2 \\
			C-4 1Mp & 997.4 & 214 & -- & $-$38.8 & 152.0 & 0.15 & 6.9 \\
			C-4 1Mp ZOA & 997.4 & 146 & -- & $-$29.6 & 133.3 & 0.13 & 0.6 \\
			
			KTB(05) MW Vir & 254 & 11 & 168 & $-$16 & 26.4 & 0.10 & 1.9 \\
			KTB(05) MW 1Mp & 956 & 16 & 168 & $-$16 & 159 & 0.17 & 3.3 \\
			MKJ(07) MW & 254 & 13 & 153.8 & $-$10.2 & 22.8 & 0.09 & 7.8 \\
			MKJ(07) And. MI & 269 & 12 & 73.4 & $-$31.5 & 45.9 & 0.17 & 1.0 \\
			MKJ(07) And. KG & 284 & 12 & 83.5 & $-$31.0 & 46.1 & 0.16 & 7.5 \\
			MKJ(09) MW & 254 & 22 & 149.6 & $-$5.3 & 28.5 & 0.11 & -- \\
			MKJ(09) And. & 589 & 23 & 60.2 & $-$30.7 & 45 & 0.08 & 15.6 \\
			Kroupa(10) MW & 254 & 24 & 156.4 & $-$2.2 & 28.9 & 0.11 & 8.2 \\
			\bottomrule
		\end{tabularx}
		
		\noindent{\footnotesize{\textsuperscript{a}~{Names of the data sets that are defined as follows: ``C-4 Vir'' and ``C-4 1Mp'': satellite data of the main galaxy with minimum stellar mass of $10^5 M_\odot$ within the virial radius and 1 Mpc radius in the ``C$-$4'' simulation presented in this study; ``C-4 Vir ZOA'' and ``C-4 1Mp ZOA'': same as ``C-4 Vir'' and ``C-4 1Mp'' except that the satellites in the ``Zone of Avoidance'' of the main galaxy are excluded in the fitting; ``KTB(05) MW Vir'' and ``KTB(05) MW 1Mp'': satellite data of the Milky Way from Table~1 of \citet{Kroupa:2005} up to a distance of 254~kpc and 956~kpc; ``MKJ(07) MW'': satellite data of the Milky Way from Table~2 of \citet{Metz:2007} with most satellites and  the algebraic least-squares(ALS) fitting method; ``MKJ(07) And. MI'' and ``MKJ(07) And. KG'': satellite data of the Andromeda galaxy from Table~2 of \citet{Metz:2007} that are originally from \citet{Koch:2006} and \citet{McConnachie:2006} with the ALS fitting method; ``MKJ(09) MW'': satellite data of the Milky Way from Table~1 of \citet{Metz:2009} plus those in \citet{Metz:2007}; ``MKJ(09) And.'': satellite data of the Andromeda galaxy from Table~2 of \mbox{\citet{Metz:2009}} plus those in \citet{Metz:2007}; ``Kroupa(10)'': satellite data of the Milky Way from Table~2 of \citet{Kroupa:2010}.} \textsuperscript{b}~{The furthest satellite's distance to the center of the main galaxy in the fitting.} \textsuperscript{c}~{Number of the satellites in the fitting.} \textsuperscript{d}~{Galactic longitude of the normal direction of the fitted DOS plane as defined in the data set.} \textsuperscript{e}~{Galactic latitude of the normal direction of the fitted DOS plane as defined in the data set.} \textsuperscript{f}~{Root mean square height of the fitted DOS plane.} \textsuperscript{g}~{Disk aspect ratio that describes the thinness of the DOS.} \textsuperscript{h}~{Distance from the fitted plane to the center of the main galaxy.}}}
	\end{table}
	
	The reason that we use different data sets for the Milky Way and  Andromeda galaxies is that the Andromeda galaxy does not have an observational  ``Zone of Avoidance'' as the Milky Way does
	. {The comparability of values obtained from the ``C-4 Vir ZOA'' and ``C-4 Vir'' data sets with those from the Milky Way and Andromeda, respectively,} is intriguing because we only have one realization of Milky-Way-like galaxy systems from the ``C'' halo initial condition and yet we get values that are similar to those from the observations. At~the very least, it shows that the high inclination angle of the fitted DOS plane in such a system is not unlikely because we only choose the ``C'' halo to have a similar mass and evolution history to the Milky Way with other factors chosen~randomly.

	The distance from the fitted plane to the center of the main galaxy is much less than the root mean square height of the fitted DOS plane in all of our data sets. Thus, the~fitted DOS planes in our study meet the consistency check~\cite{Metz:2007} that constituent satellites orbit within the host potential such that a disk made up of a virialized satellite population passes near the origin (or host galaxy). Disks with distances greater than one disk height from the host are considered unphysical.
	We measure ``$\Delta/R_{\rm {cut}}$'', the~flattening parameter defined in~\cite{Kroupa:2005}, where $\Delta$ is the rms height of the disk and $R_{\rm {cut}}$ is the farthest distance to the Galactic Centre in the satellite sample. This disk aspect ratio describes the thinness of the DOSs in our study{. I}t is generally larger ($\sim$10--30\%) than the ones from observation ($\sim$10--20\%). This means that the fitted DOSs are thicker using the data from our simulations. 
	However, when considering a scale of 1 Mpc, the~fitted DOS appears thinner. This is because, at~this larger scale, the~satellite distributions align more closely with the shape of the local dark matter~distribution.

	\subsection{Dynamical~Properties}
	
	From  observational data on the proper motion of the satellites, it has been argued that some of the Milky Way's satellites may have coherent motion in phase space.  This coherence implies that they may share a common origin \citep{Lynden-Bell:1995,Kroupa:2005,Metz:2007,Metz:2008}. While some numerical studies have shown \citep{Libeskind:2007,Libeskind:2009,Lovell:2011,Deason:2011}  {that} the direction of the satellites' angular momentum can align closely with that of the main galaxy, it remains challenging \citep{Libeskind:2009,Deason:2011} to identify a satellite system in the simulations that matches the orbiting pole alignment observed in~\cite{Metz:2008}.
	
\begin{figure}[H]
		\includegraphics[angle=0, width=3.5in]{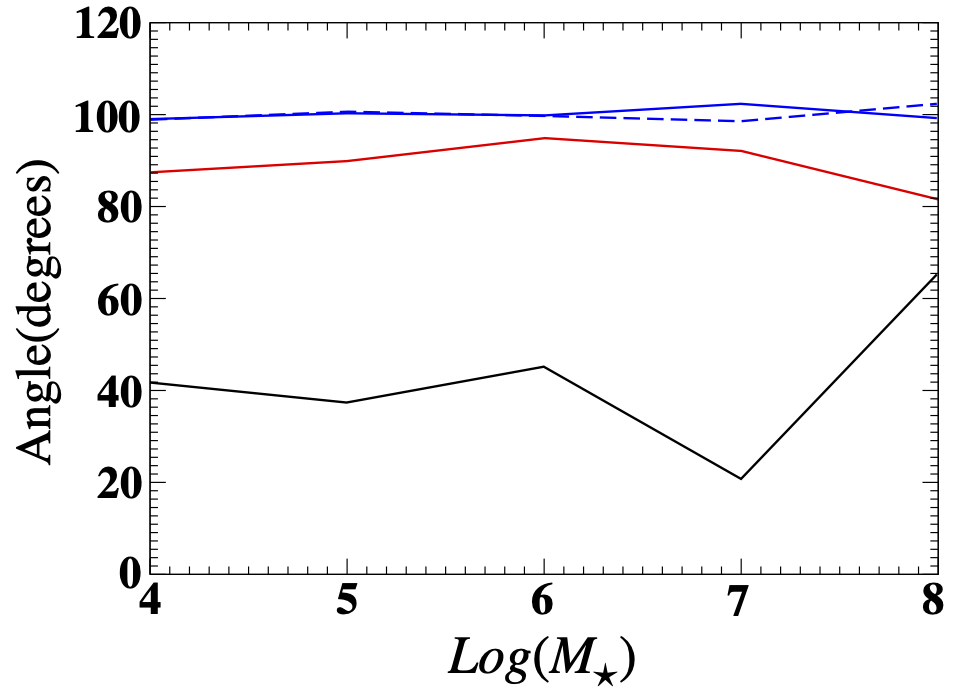}
		\caption{Angular relationships relative to the direction of the angular momentum of the DOS as a function of minimum stellar mass in the satellites within the virial radius in the ``C$-$4'' simulation at redshift z = 0. 
			The line in solid blue represents the angle between the directions of the angular momenta of the DOS and the stellar disk of the main galaxy. The~line in dashed blue represents the angle between the directions of the angular momenta of the DOS and the host dark matter halo.
			The {red line} represents the angle between the angular momentum vector of the DOS and the normal direction of the velocity dispersion vector, denoted as ${\vec \sigma}_{\rm DOS}$, of~the DOS.
			The {black line} represents the angle between the normal direction of the DOS, represented by ${\vec n}_{\rm DOS}$, and~the direction of the angular momentum vector of the DOS.
		}
	\label{fig_dn_da_vir}
\end{figure}

	The bottom {black solid} line on Figure~\ref{fig_dn_da_vir} 
	{represents the angular separation between the normal direction of the DOS and the angular momentum vector direction of the DOS, plotted} as a function of the minimum stellar mass of the satellites within the virial radius in the  C$-$4 simulation at redshift $z=0$.  If~the satellites within the fitted DOS are rotationally supported, these two vectors should closely, if~not perfectly, align. However, the~angular separation between these two vectors ranges between  $20 ^{\circ}$  and $70 ^{\circ}$ showing no discernible  trend with satellite mass, {and} the angle never approaches zero. This suggests that, at~least within the context of the DOS fitted with satellites in this study's simulation, the~satellites do not exhibit coherent rotation within the DOS~plane.

	The middle   line in Figure~\ref{fig_dn_da_vir} further supports this observation.  This shows the angle between the direction of the angular momentum with respect to the center of the main galaxy and the normal direction of the velocity dispersion $\vec {\sigma}_{\rm DOS}$ of the DOS. This is plotted as a function of minimum stellar mass in the satellites while keeping other parameters the same. Given  {that} the normal direction of the velocity dispersion represents the direction of the least bulk velocity flow of the satellites, the very large angle of about $90 ^{\circ}$ in \mbox{Figure~\ref{fig_dn_da_vir}} indicates a substantial bulk velocity flow along a direction parallel to the angular momentum {vector} $\mathbf{L}_{\rm Main}$ of the main galaxy. This suggests that the bulk motion of the fitted DOS in this simulation is characteristic of an infalling system, rather than an orbiting one. For~an orbiting motion, the~bulk motion would point parallel to the plane of the DOS, not perpendicular to it. Thus, there is no indication that the fitted DOS plane has a common orbital plane with the satellites within it.
	
	To examine the relation between the directions of the angular momenta of the DOS and the main galaxy, {in the top {solid} line on Figure~\ref{fig_dn_da_vir}, we plot the angle {(solid line)} between the direction of the angular momentum vectors of the DOS and that of the stellar disk of the main galaxy ${\bf L}_{\rm main}$ as a function of minimum stellar mass in the satellites.  The dashed line shows} the angle with respect to the host dark matter halo.
	There are several distinct features in this figure: First, the~angle between the directions of the angular momentum of the DOS and  the plane of the stellar disk of the main galaxy or the host dark matter halo is independent of the stellar mass in the satellites. Second, the~main galaxy and the host dark matter halo have almost the same angular momentum direction. Most importantly, one  can see in Figure~\ref{fig_dn_da_vir} that the angle between the directions of the angular momentum of the DOS, the~stellar disk of the main galaxy and the host dark matter halo is always about $100 ^{\circ}$. This indicates that the direction of the angular momentum of the DOS with respect to the center of the main galaxy is almost perpendicular to that of the main galaxy {\it and} the DM sub-halos.
	This shows a very distinct difference in the dynamics between the satellites and the main galaxy. The~dynamics of the satellites reflect the dynamics of infall toward the main disk. This differs from a rotationally supported system orbiting around the plane of the~DOS.
\unskip

\subsection{Time Evolution}

Until now, we have been concerned with the properties of the satellite systems  {during} the current epoch at redshift $z=0$.  This is because most of the observational data of the satellites are from the current satellite systems of the Milky Way and Andromeda galaxy. 
{More recent observations of extragalactic disk galaxy systems in data from the 2dF Galaxy Redshift Survey and SDSS have extended this analysis to higher redshift {s} to probe deviations of the satellite spatial distribution and orientation from those of the Milky Way and Andromeda systems~\cite{Donoso:2006,Wang:2010}. 
	Therefore, it is also essential} to study the time evolution of the satellite systems around the main galaxy if one  {wishes to understand} the large-scale dynamics of such systems.

\begin{figure}[H]
	\includegraphics[trim=0cm 0cm 0cm 0cm, clip=true, angle=0, width=0.3\textwidth]{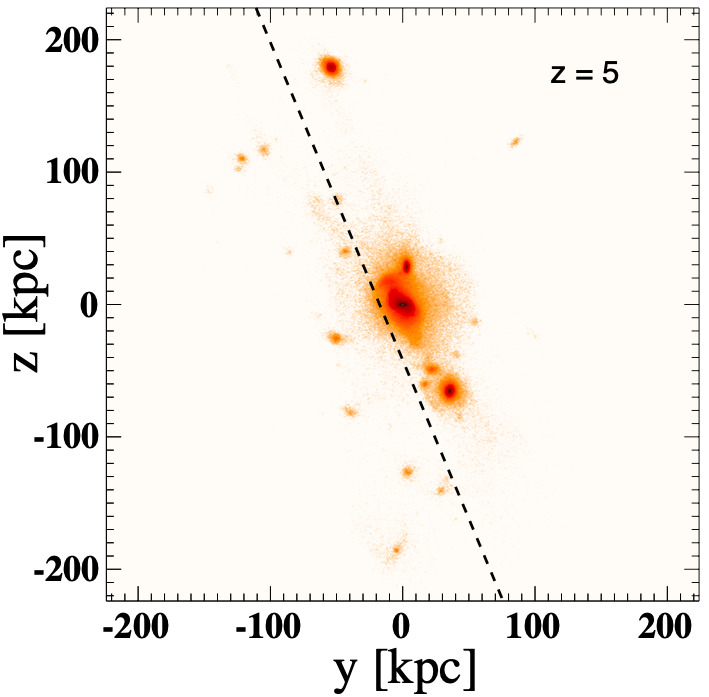}
	\hspace{0cm}
	\includegraphics[trim=0cm 0cm 0cm 0cm, clip=true, angle=0, width=0.3\textwidth]{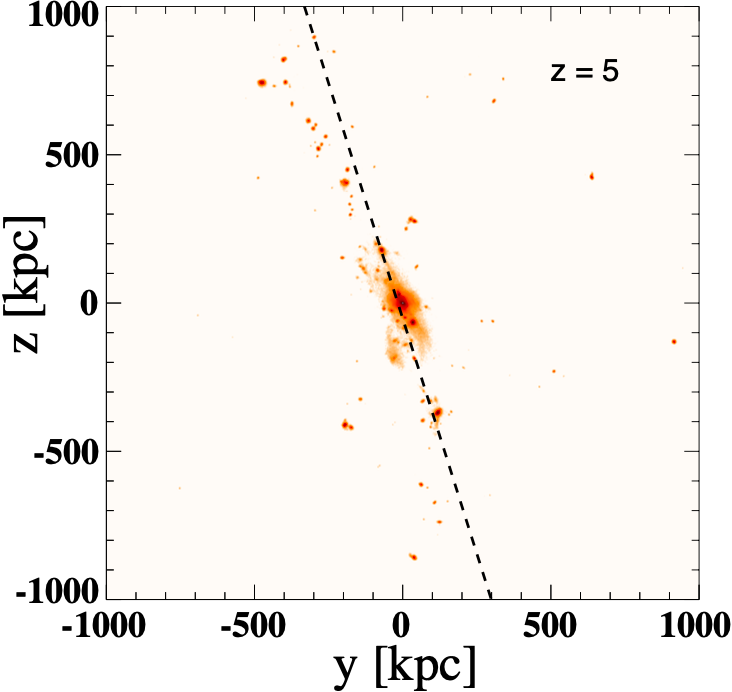}
	\hspace{0cm}
	\includegraphics[trim=0cm 0cm 0cm 0cm, clip=true, angle=0, width=0.3\textwidth]{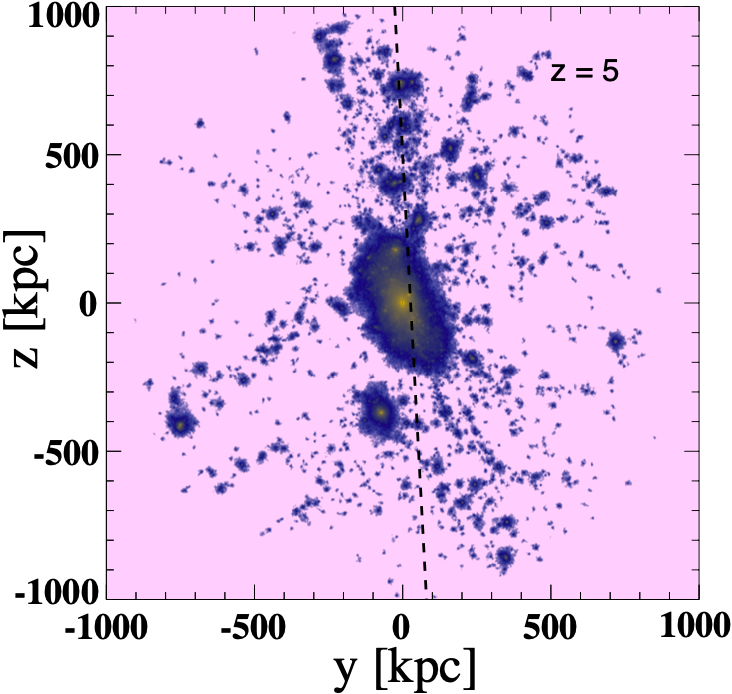} \\
	\vspace{0cm}
	\includegraphics[trim=0cm 0cm 0cm 0cm, clip=true, angle=0, width=0.3\textwidth]{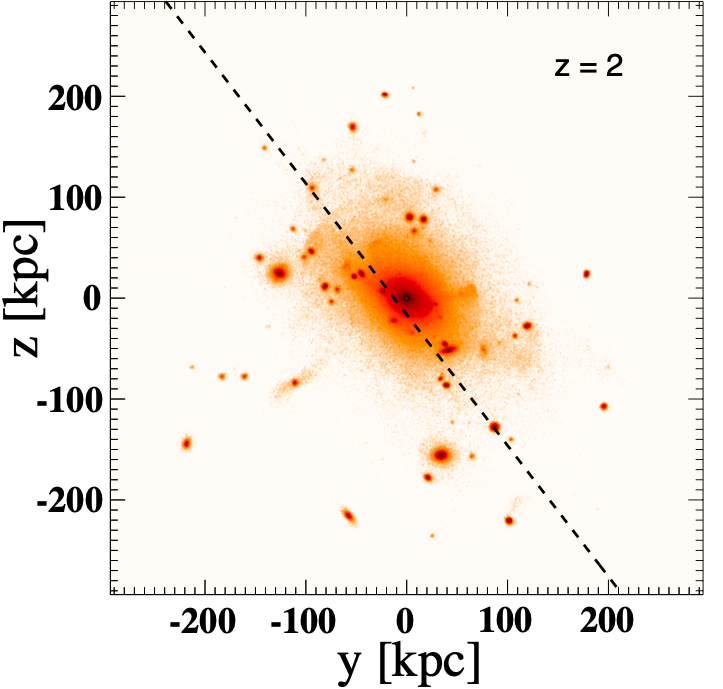}
	\hspace{0cm}
	\includegraphics[trim=0cm 0cm 0cm 0cm, clip=true, angle=0, width=0.3\textwidth]{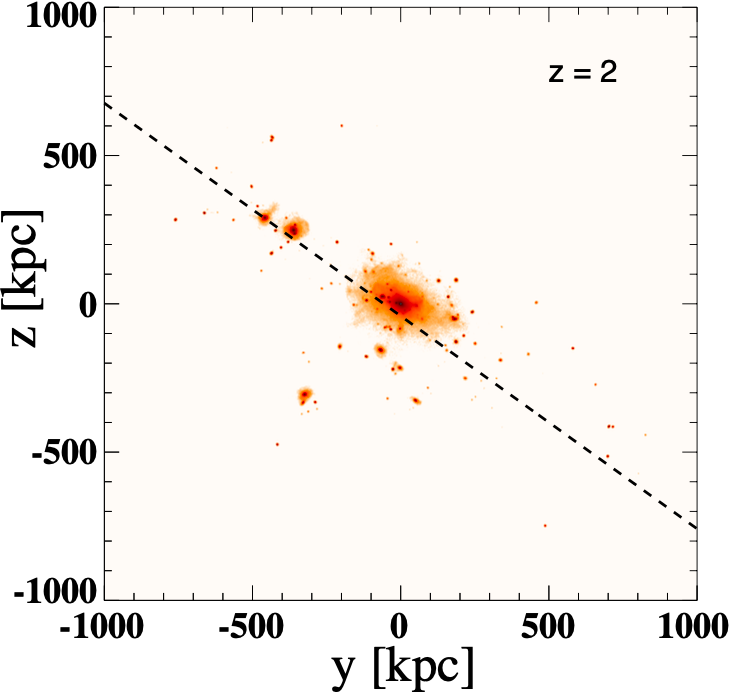}
	\hspace{0cm}
	\includegraphics[trim=0cm 0cm 0cm 0cm, clip=true, angle=0, width=0.3\textwidth]{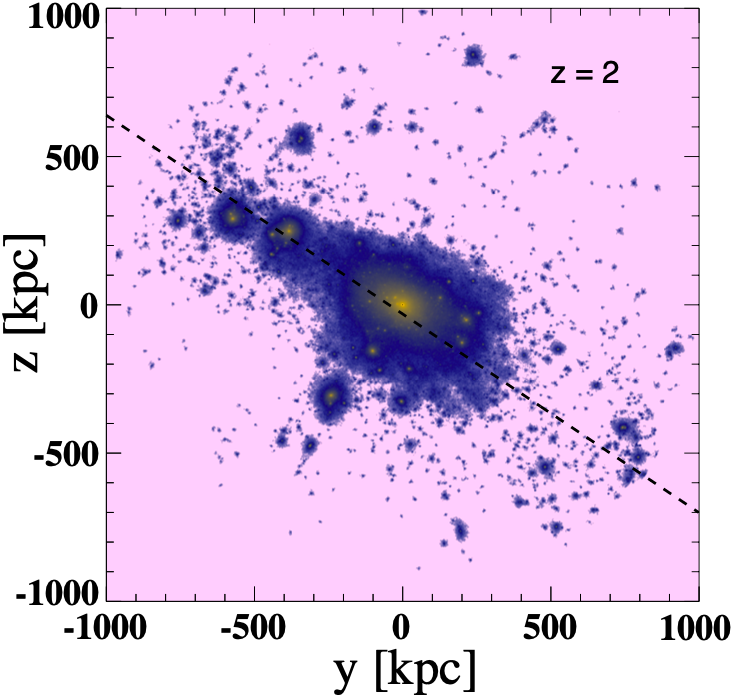} \\
	\vspace{0cm}
	\includegraphics[trim=0cm 0cm 0cm 0cm, clip=true, angle=0, width=0.3\textwidth]{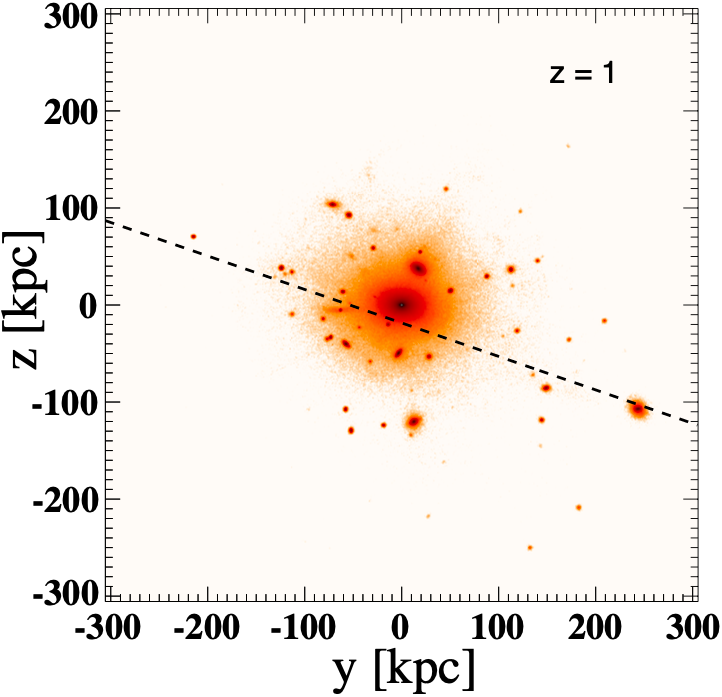}
	\hspace{0cm}
	\includegraphics[trim=0cm 0cm 0cm 0cm, clip=true, angle=0, width=0.3\textwidth]{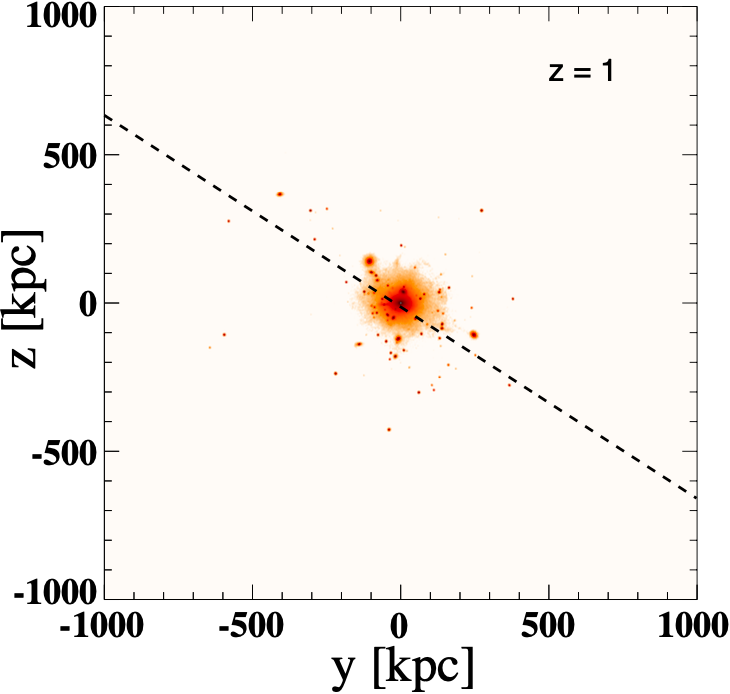}
	\hspace{0cm}
	\includegraphics[trim=0cm 0cm 0cm 0cm, clip=true, angle=0, width=0.3\textwidth]{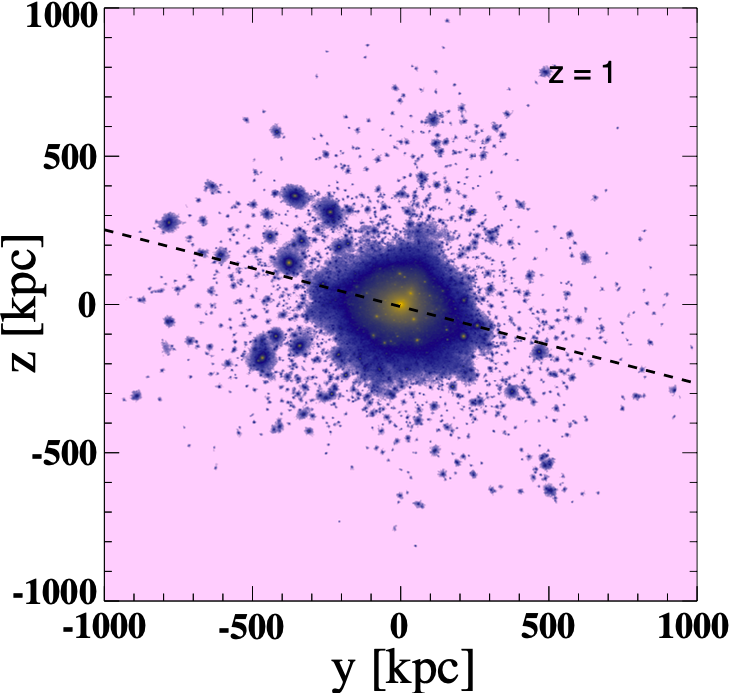} \\
	\vspace{0cm}
	\includegraphics[trim=0cm 0cm 0cm 0cm, clip=true, angle=0, width=0.3\textwidth]{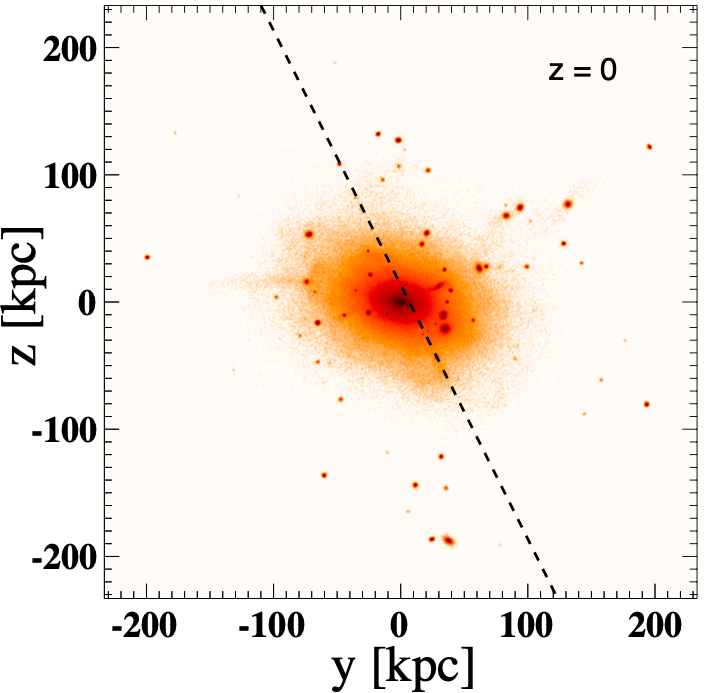}
	\hspace{0cm}
	\includegraphics[trim=0cm 0cm 0cm 0cm, clip=true, angle=0, width=0.3\textwidth]{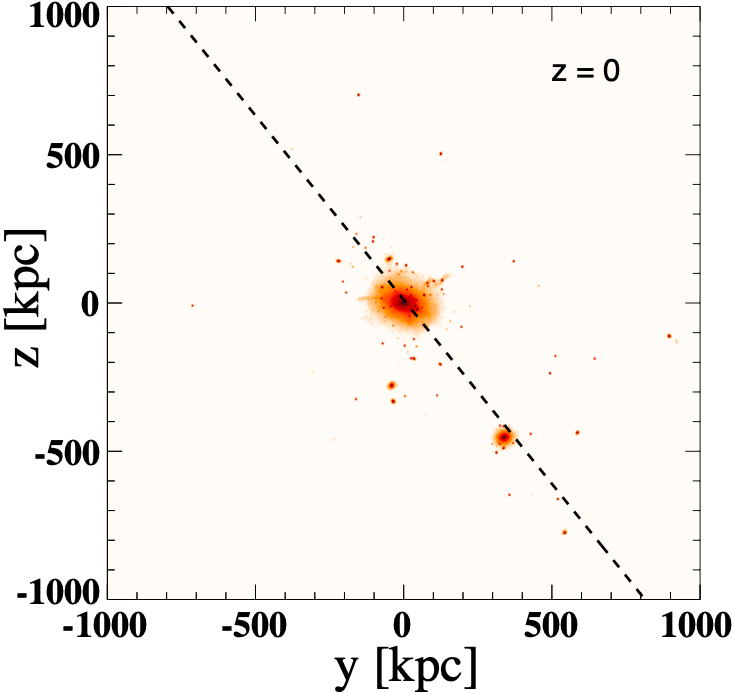}
	\hspace{0cm}
	\includegraphics[trim=0cm 0cm 0cm 0cm, clip=true, angle=0, width=0.3\textwidth]{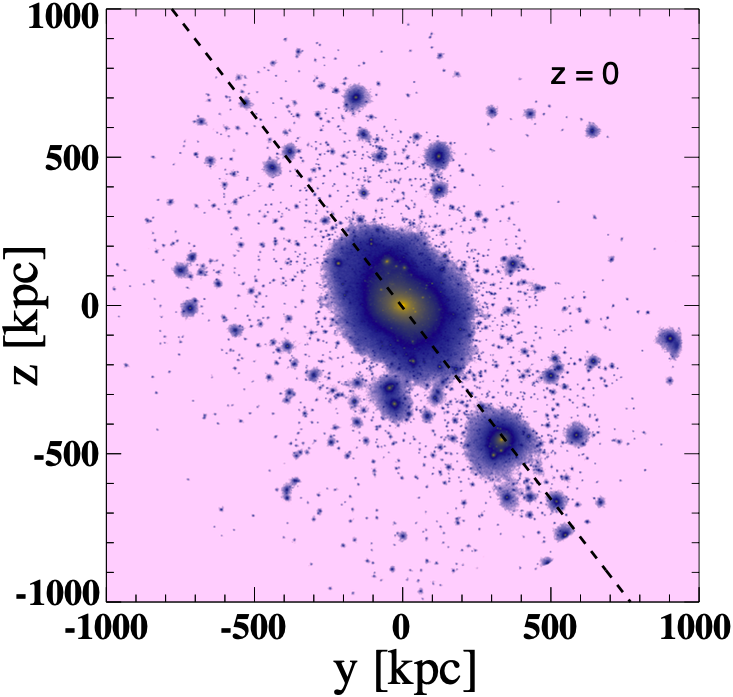}
	\vspace{0cm}
	\caption{{The} edge-on views of the main galaxy's stellar disk and the DOS formed by the satellites with {a} minimum stellar mass of $10^5 M_\odot$ within the virial radius (left column) and 1 Mpc radius (middle column) and all dark matter sub-halos within 1 Mpc radius (right column) in the ``C$-$4'' simulation at various redshifts.}
	\label{fig_dos_1mp_z}
\end{figure}

In Figure~\ref{fig_dos_1mp_z}, we render the edge-on views of the main galaxy's stellar disk and the DOS formed by the luminous satellites within the virial radius (left column) and a 1 Mpc radius (middle column) and the DOS formed by all dark matter sub-halos within a 1 Mpc radius (right column) from redshift $z=5$ to $z=0$. From~this figure one can clearly see that the DOS formed by the satellites on different scales and the DOS formed by the DM sub-halos have similar orientations relative to the stellar disk of the main galaxy at each~redshift. 

\begin{figure}[H]
	\includegraphics[angle=0, width=3.5in]{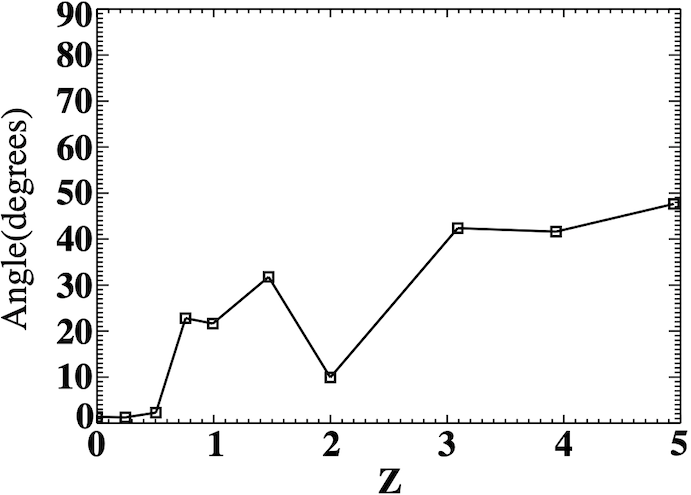}
	\caption{The angle between the normal directions of the DOS ${\vec n}_{\rm DOS}$ formed by the satellites with minimum stellar mass of $10^5 M_\odot$ and all dark matter sub-halos within 1 Mpc radius as a function of redshift $z$ in the ``C$-$4'' simulation.}
	\label{fig_dn_05_00_1mp_z}
\end{figure}

This is especially true at lower redshifts {($z < 3$)} as is manifested in Figure~\ref{fig_dn_05_00_1mp_z} in which we plot the angle between the normal directions of the DOS formed by the luminous satellites and that by all dark matter sub-halos within a 1 Mpc radius as a function of redshift $z$. As~shown in Figure~\ref{fig_dn_05_00_1mp_z}, the~angle gradually decreases from about $50 ^{\circ}$ at $z=5$ to almost zero at $z=0$. This strongly suggests that the spatial distribution of the luminous satellites traces that of the dark matter on a 1 Mpc scale at low redshifts {($z < 3$)}. Before~$z=0$, the~system has had sufficient time to evolve and settle into a more stable configuration. The~abrupt change in the angle at $z=2$ reflects the fact that a major accretion event happened around that time. It appears that after a merger event starts, the~angle increases rapidly due to the complex dynamics induced by ongoing mergers, causing disruption and deviations from the previously formed alignment. As~time progresses, the~system undergoes a transitional period during which the combined influence of ongoing disruptions and the emergence of a more coherent structure might contribute to the observed variability in the angle. Eventually, around $z=0.7$, as~the system settles into a more stabilized configuration, the~dynamics become dominated by the dark matter component, and~the angle between the normal directions declines rapidly. This decline indicates a restoration of alignment between the satellite distribution and the underlying dark matter distribution. This trend of alignment near $z=0$ suggests a compelling association between the dynamical evolution of satellite galaxies and the underlying DM sub-halo~structure. 

\unskip

\unskip

\section{Summary and~Conclusions}\label{sec4}
In summary, we utilized original high-resolution hydrodynamical simulations to examine the satellite system surrounding a Milky-Way-like galaxy. The~incorporated baryonic physics in these original simulations enabled a comprehensive study of the spatial and dynamical properties of luminous satellites in such a system. The~high mass resolution ($\sim$$3\times 10^4 M_\odot$ for star particles) in the simulations makes the capture of the formation of low luminosity satellites with $M > 10^5 M_\odot$ possible in our study. Confining our study to the galaxy with the most Milky-Way like well-formed disk, we found, intriguingly, the~presence of a DOS. While the existence of a DOS in a single galaxy does not address its likelihood of occurrence, it {allowed} an opportunity to investigate the properties and dynamics of the simulated~DOS. 

Our conclusions from this study are as~follows:
\begin{enumerate}
	\item   We find that the properties of the fitted DOS planes are almost independent of the lower mass thresholds (proxy for luminosity) applied to the satellite samples, from~$10^5 M_\odot$ to $10^8 M_\odot$. 
	\item  The DOS formed by the luminous satellites and by the dark matter sub-halos have similar orientations. {That is{,} the orientation of the DM DOS of Figure~\ref{fig_dm_dos_vir} and the most similar stellar DOS of Figure~\ref{fig_dos_vir} are almost identical.}
	\item Comparing our results with some previous observational studies, we find that the fitted DOS plane in our simulations can exhibit a significant inclination angle relative to the galactic disk  {which is slightly larger than the one observed} in the Milky Way. However, the~fitted DOS disk in the simulations is also thicker than that of the Milky Way. {That is, the~DOS aspect ratios given in Table~\ref{tab_dos_par} for the simulations range from 0.1 to 0.3, whereas the observed aspect ratio for the Milky Way DOS is $\sim$0.1--0.2.}
	\item We also find that the ``Zone of Avoidance'' of the Milky Way may have some impacts on the fitted parameters. Indeed, when including the ZOA in the simulations, the~properties (angles, distances, etc.) from the simulations more closely resemble those from observations.
	\item Perhaps the most important conclusion from this study is the pronounced difference in the direction of the angular momentum of satellites in the DOS plane compared to the main galaxy. {In the simulations, they are nearly perpendicular to each other suggesting that the satellite motion is consistent with infall from a larger scale. This {is consistent with} the observational data, indicating that, at least in our current study, the DOS is not rotationally supported.
		\item Our analysis shows that the anisotropic distribution of the satellites in the simulations can have characteristic properties such as angles and distances similar to the characteristics of the Milky Way's fitted DOS from observations. }
\end{enumerate}

Based upon these conclusions we speculate that the DOS derived from the positions of the nearby satellites within the main galaxy's virial radius may reflect the distribution of the satellites and DM sub-halos on a larger scale. 
Also, we suggest  that  the fitted DOS plane has been greatly affected by major accretion processes that happened in the evolution history of the main galaxy. The~newly formed more luminous satellites accreted in these processes may have dominant roles in the formation of the observed DOS. 

What is needed are  much larger samples of higher-resolution realistic simulations of Milky-Way-like galaxies to substantiate that the observed DOS can be naturally formed in a standard $\Lambda CDM$ cosmology \citep{Sawala2022}. We will address this in a future work (cf.~\cite{Gonzalez23}). Further study of the DOS will also benefit from recent and forthcoming observational efforts including the potential discovery of new satellites (e.g., LSST), the~measurements of the proper motions of satellites (e.g., Gaia~\cite{Gaia:2018}) and their line-of-sight velocities (e.g., DESI Milky Way Survey,~\cite{Cooper:2023}). The~latter effort will also lead to the observation of extended features around sixteen MW satellites in the survey footprint~\cite{Cooper:2023}.

\vspace{6pt} 
\authorcontributions{Conceptualization, X.Z. and G.J.M.; methodology, X.Z.; software, X.Z.; validation, X.Z.; formal analysis, X.Z., G.J.M. and G.T.; investigation, X.Z., G.J.M., G.T. and L.A.P.; resources, X.Z. and G.J.M.; data curation, X.Z. and G.T.; writing---original draft preparation, X.Z. and G.J.M.; writing---review and editing, X.Z., G.J.M., G.T. and L.A.P.; visualization, X.Z., G.J.M. and G.T.; supervision, X.Z. and G.J.M.; project administration, X.Z. and G.J.M.; funding acquisition, X.Z. and G.J.M. All authors have read and agreed to the published version of the manuscript.}


\funding{~\textls[-15]{This research was funded by ~the National Science Foundation under grants AST-0965694 and AST-1009867 and ~the U.S. Department of Energy under Nuclear Theory Grant DE-FG02-95-ER40934.}~ }

\dataavailability{Data supporting reported results can be obtained by request from the~authors.}

\acknowledgments {We acknowledge the Research Computing and Cyber-infrastructure unit of Information Technology Services at the Pennsylvania State University for providing computational resources and services that have contributed to the early phases of the research reported in this paper (\url{https://researchcomputing.psu.edu})  last accessed on 15 May 2014.}

\conflictsofinterest{The authors declare no conflict of interest.}

\begin{adjustwidth}{-\extralength}{0cm}
\printendnotes[custom] 

\reftitle{References}

\PublishersNote{}
\end{adjustwidth}


\begin{thebibliography}{999}

\bibitem[{Lynden-Bell}(1976)]{Lynden-Bell:1976}
{Lynden-Bell}, D.
\newblock {Dwarf galaxies and globular clusters in high velocity hydrogen
streams}.
\newblock {\em Mon. Not. R. Astron. Soc.} {\bf 1976}, {\em 174},~695--710. [\href{http://doi.org/10.1093/mnras/174.3.695}{CrossRef}]

\bibitem[{Kunkel} and {Demers}(1976)]{Kunkel:1976}
{Kunkel}, W.E.; {Demers}, S.
\newblock The Magellanic Plane. In {\em Royal Greenwich Observatory Bulletin, Proceedings of the Galaxy and the Local Group, Tercentenary Symposium, {Hailsham, UK, 22--25 July 1975}};
\newblock {Dickens, R.J., Perry, J.E., Smith, F.G., King, I.R.}, Eds.; Royal Greenwich Observatory: {Herstmonceux, UK}, 1976; Volume 182, p. 241.

\bibitem[{Kroupa} {et~al.}(2005){Kroupa}, {Theis}, and {Boily}]{Kroupa:2005}
{Kroupa}, P.; {Theis}, C.; {Boily}, C.M.
\newblock {The great disk of Milky-Way satellites and cosmological
sub-structures}.
\newblock {\em Astron. Astrophys.} {\bf 2005}, {\em 431},~517--521.
 [\href{http://dx.doi.org/10.1051/0004-6361:20041122}{CrossRef}]

\bibitem[{Kroupa} {et~al.}(2010){Kroupa}, {Famaey}, {de Boer},
{Dabringhausen}, {Pawlowski}, {Boily}, {Jerjen}, {Forbes}, {Hensler}, and
{Metz}]{Kroupa:2010}
{Kroupa}, P.; {Famaey}, B.; {de Boer}, K.S.; {Dabringhausen}, J.; {Pawlowski},
M.S.; {Boily}, C.M.; {Jerjen}, H.; {Forbes}, D.; {Hensler}, G.; {Metz}, M.
\newblock {Local-Group tests of dark-matter concordance cosmology. Towards a
new paradigm for structure formation}.
\newblock {\em Astron. Astrophys.} {\bf 2010}, {\em 523},~A32.
 [\href{http://dx.doi.org/10.1051/0004-6361/201014892}{CrossRef}]

\bibitem[{Kroupa}(2012)]{Kroupa:2012}
{Kroupa}, P.
\newblock {The Dark Matter Crisis: Falsification of the Current Standard Model
of Cosmology}.
\newblock {\em Publ. Astron. Soc. Aust.} {\bf 2012}, {\em 29},~395--433.
 [\href{http://dx.doi.org/10.1071/AS12005}{CrossRef}]

\bibitem[{Metz} {et~al.}(2009){Metz}, {Kroupa}, and {Jerjen}]{Metz:2009}
{Metz}, M.; {Kroupa}, P.; {Jerjen}, H.
\newblock {Discs of satellites: The new dwarf spheroidals}.
\newblock {\em Mon. Not. R. Astron. Soc.} {\bf 2009}, {\em 394},~2223--2228.
 [\href{http://dx.doi.org/10.1111/j.1365-2966.2009.14489.x}{CrossRef}]

\bibitem[{Simon}(2019)]{Simon:2019}
{Simon}, J.D.
\newblock {The Faintest Dwarf Galaxies}.
\newblock {\em Annu. Rev. Astron. Astrophys.} {\bf 2019}, {\em 57},~375--415.
[\href{http://dx.doi.org/10.1146/annurev-astro-091918-104453}{CrossRef}]

\bibitem[{Grebel} {et~al.}(1999){Grebel}, {Kolatt}, and
{Brandner}]{Grebel:1999}
{Grebel}, E.K.; {Kolatt}, T.; {Brandner}, W.
\newblock {Orbits versus Star Formation Histories: A Progress Report}.
\newblock In \emph{The Stellar Content of Local Group Galaxies, Proceedings of the {192nd Symposium of the International Astronomical Union, Cape Town, South Africa, 7--11 September 1998}};
{Whitelock, P., Cannon, R.}, Eds.; {Astronomical Society of the Pacific (ASP): San Francisco, CA, USA}, 1999; Volume 192, p. 447.

\bibitem[{Hartwick}(2000)]{Hartwick:2000}
{Hartwick}, F.D.A.
\newblock {The Structure of the Outer Halo of the Galaxy and its Relationship
to Nearby Large-Scale Structure}.
\newblock {\em Astron. J.} {\bf 2000}, {\em 119},~2248--2253.
 [\href{http://dx.doi.org/10.1086/301332}{CrossRef}]

\bibitem[{Koch} and {Grebel}(2006)]{Koch:2006}
{Koch}, A.; {Grebel}, E.K.
\newblock {The Anisotropic Distribution of M31 Satellite Galaxies: A Polar
Great Plane of Early type Companions}.
\newblock {\em Astron. J.} {\bf 2006}, {\em 131},~1405--1415.
 [\href{http://dx.doi.org/10.1086/499534}{CrossRef}]

\bibitem[{McConnachie} and {Irwin}(2006)]{McConnachie:2006}
{McConnachie}, A.W.; {Irwin}, M.J.
\newblock {The satellite distribution of M31}.
\newblock {\em Mon. Not. R. Astron. Soc.} {\bf 2006}, {\em 365},~902--914.
 [\href{http://dx.doi.org/10.1111/j.1365-2966.2005.09771.x}{CrossRef}]

\bibitem[{Metz} {et~al.}(2007){Metz}, {Kroupa}, and {Jerjen}]{Metz:2007}
{Metz}, M.; {Kroupa}, P.; {Jerjen}, H.
\newblock {The spatial distribution of the Milky Way and Andromeda satellite
galaxies}.
\newblock {\em Mon. Not. R. Astron. Soc.} {\bf 2007}, {\em 374},~1125--1145.
 [\href{http://dx.doi.org/10.1111/j.1365-2966.2006.11228.x}{CrossRef}]

\bibitem[Martin {et~al.}(2009)Martin, McConnachie, Irwin, Widrow, Ferguson,
Ibata, Dubinski, Babul, Chapman, Fardal, Lewis, Navarro, and
Rich]{Martin:2009}
Martin, N.F.; McConnachie, A.W.; Irwin, M.; Widrow, L.M.; Ferguson, A.M.N.;
Ibata, R.A.; Dubinski, J.; Babul, A.; Chapman, S.; Fardal, M.;  et~al.
\newblock {PAndAS}' {Cubs}: {Discovery} {of} {Two} {New} {Dwarf} {Galaxies}
{in} {the} {Surroundings} {of} {the} {Andromeda} {and} {Triangulum}
{Galaxies}.
\newblock {\em  Astrophys. J.} {\bf 2009}, {\em 705},~758--765.
[\href{http://dx.doi.org/10.1088/0004-637X/705/1/758}{CrossRef}]

\bibitem[Richardson {et~al.}(2011)Richardson, Irwin, McConnachie, Martin,
Dotter, Ferguson, Ibata, Chapman, Lewis, Tanvir, and Rich]{Richardson:2011}
Richardson, J.C.; Irwin, M.J.; McConnachie, A.W.; Martin, N.F.; Dotter, A.L.;
Ferguson, A.M.N.; Ibata, R.A.; Chapman, S.C.; Lewis, G.F.; Tanvir, N.R.;
et~al.
\newblock {PAndAS}' {Progeny}: {Extending} {the} M31 {Dwarf} {Galaxy} {Cabal}.
\newblock {\em  Astrophys. J.} {\bf 2011}, {\em 732},~76.
 [\href{http://dx.doi.org/10.1088/0004-637X/732/2/76}{CrossRef}]

\bibitem[Martin {et~al.}(2013{\natexlab{a}})Martin, Slater, Schlafly,
Morganson, Rix, Bell, Laevens, Bernard, Ferguson, Finkbeiner, Burgett,
Chambers, Hodapp, Kaiser, Kudritzki, Magnier, Morgan, Price, Tonry, and
Wainscoat]{Martin:2013a}
{Martin, N.F.; Slater, C.T.; Schlafly, E.F.; Morganson, E.; Rix, H.W.; Bell,
E.F.; Laevens, B.P.M.; Bernard, E.J.; Ferguson, A.M.N.; Finkbeiner, D.P.;
et~al.}
\newblock {Lacerta} I {and} {Cassiopeia} {III}. {Two} {Luminous} {and}
{Distant} {Andromeda} {Satellite} {Dwarf} {Galaxies} {Found} {in} {the}
3$\uppi$ {Pan}-{Starrs}1 {Survey}.
\newblock {\em  Astrophys. J.} {\bf 2013}, {\em 772},~15.
[\href{http://dx.doi.org/10.1088/0004-637X/772/1/15}{CrossRef}]


\bibitem[Martínez-Delgado {et~al.}(2021)Martínez-Delgado, Karim, Charles,
Boschin, Monelli, Collins, Donatiello, and Alfaro]{Martinez:2022}
Martínez-Delgado, D.; Karim, N.; Charles, E.J.E.; Boschin, W.; Monelli, M.;
Collins, M.L.M.; Donatiello, G.; Alfaro, E.J.
\newblock {Pisces VII: Discovery of a possible satellite of Messier 33 in the
DESI legacy imaging surveys}.
\newblock {\em Mon. Not. R. Astron. Soc.} {\bf 2021},
{\em 509},~16--24.
[\href{http://dx.doi.org/10.1093/mnras/stab2797}{CrossRef}]

\bibitem[{Lynden-Bell} and {Lynden-Bell}(1995)]{Lynden-Bell:1995}
{Lynden-Bell}, D.; {Lynden-Bell}, R.M.
\newblock {Ghostly streams from the formation of the Galaxy's halo}.
\newblock {\em Mon. Not. R. Astron. Soc.} {\bf 1995}, {\em 275},~429--442. [\href{http://dx.doi.org/10.1093/mnras/275.2.429}{CrossRef}]

\bibitem[{Metz} {et~al.}(2008){Metz}, {Kroupa}, and {Libeskind}]{Metz:2008}
{Metz}, M.; {Kroupa}, P.; {Libeskind}, N.I.
\newblock {The Orbital Poles of Milky Way Satellite Galaxies: A Rotationally
Supported Disk of Satellites}.
\newblock {\em Astrophys. J.} {\bf 2008}, {\em 680},~287--294.
[\href{http://dx.doi.org/10.1086/587833}{CrossRef}]

\bibitem[{Santos-Santos} {et~al.}(2020){Santos-Santos},
{Dom{\'\i}nguez-Tenreiro}, {Artal}, {Pedrosa}, {Bignone},
{Mart{\'\i}nez-Serrano}, {G{\'o}mez-Flechoso}, {Tissera}, and
{Serna}]{Santos-Santos:2020a}
{Santos-Santos}, I.; {Dom{\'\i}nguez-Tenreiro}, R.; {Artal}, H.; {Pedrosa},
S.E.; {Bignone}, L.; {Mart{\'\i}nez-Serrano}, F.; {G{\'o}mez-Flechoso},
M.{\'A}.; {Tissera}, P.B.; {Serna}, A.
\newblock {Planes of Satellites around Simulated Disk Galaxies. I. Finding
High-quality Planar Configurations from Positional Information and Their
Comparison to MW/M31 Data}.
\newblock {\em Astrophys. J.} {\bf 2020}, {\em 897},~71.
 [\href{http://dx.doi.org/10.3847/1538-4357/ab7f29}{CrossRef}]

\bibitem[{Wang} {et~al.}(2020){Wang}, {Libeskind}, {Tempel}, {Pawlowski},
{Kang}, and {Guo}]{Wang:2020}
{Wang}, P.; {Libeskind}, N.I.; {Tempel}, E.; {Pawlowski}, M.S.; {Kang}, X.;
{Guo}, Q.
\newblock {The Alignment of Satellite Systems with Cosmic Filaments in the SDSS
DR12}.
\newblock {\em Astrophys. J.} {\bf 2020}, {\em 900},~129.
 [\href{http://dx.doi.org/10.3847/1538-4357/aba6ea}{CrossRef}]

\bibitem[{Pawlowski} and {Tony Sohn}(2021)]{Pawlowski:2021}
{Pawlowski}, M.S.; {Tony Sohn}, S.
\newblock {On the Co-orbitation of Satellite Galaxies along the Great Plane of
Andromeda: NGC 147, NGC 185, and Expectations from Cosmological Simulations}.
\newblock {\em Astrophys. J.} {\bf 2021}, {\em 923},~42.
 [\href{http://dx.doi.org/10.3847/1538-4357/ac2aa9}{CrossRef}]

\bibitem[{Keller} {et~al.}(2012){Keller}, {Mackey}, and {Da
Costa}]{Keller:2012}
{Keller}, S.C.; {Mackey}, D.; {Da Costa}, G.S.
\newblock {The Globular Cluster System of the Milky Way: Accretion in a
Cosmological Context}.
\newblock {\em Astrophys. J.} {\bf 2012}, {\em 744},~57.
 [\href{http://dx.doi.org/10.1088/0004-637X/744/1/57}{CrossRef}]

\bibitem[{Pawlowski} {et~al.}(2012){Pawlowski}, {Pflamm-Altenburg}, and
{Kroupa}]{Pawlowski:2012}
{Pawlowski}, M.S.; {Pflamm-Altenburg}, J.; {Kroupa}, P.
\newblock {The VPOS: A vast polar structure of satellite galaxies, globular
clusters and streams around the Milky Way}.
\newblock {\em Mon. Not. R. Astron. Soc.} {\bf 2012}, {\em 423},~1109--1126.
 [\href{http://dx.doi.org/10.1111/j.1365-2966.2012.20937.x}{CrossRef}]

\bibitem[{Cautun} {et~al.}(2015){Cautun}, {Wang}, {Frenk}, and
{Sawala}]{Cautun2015}
{Cautun}, M.; {Wang}, W.; {Frenk}, C.S.; {Sawala}, T.
\newblock {A new spin on discs of satellite galaxies}.
\newblock {\em Mon. Not. R. Astron. Soc.} {\bf 2015}, {\em 449},~2576--2587.
 [\href{http://dx.doi.org/10.1093/mnras/stv490}{CrossRef}]

\bibitem[{Libeskind} {et~al.}(2016){Libeskind}, {Guo}, {Tempel}, and
{Ibata}]{Libeskind2016}
{Libeskind}, N.I.; {Guo}, Q.; {Tempel}, E.; {Ibata}, R.
\newblock {The Lopsided Distribution of Satellite Galaxies}.
\newblock {\em Astrophys. J.} {\bf 2016}, {\em 830},~121.
 [\href{http://dx.doi.org/10.3847/0004-637X/830/2/121}{CrossRef}]

\bibitem[{Maji} {et~al.}(2017){Maji}, {Zhu}, {Marinacci}, and
{Li}]{Maji2017}
{Maji}, M.; {Zhu}, Q.; {Marinacci}, F.; {Li}, Y.
\newblock {Is There a Disk of Satellites around the Milky Way?}
\newblock {\em Astrophys. J.} {\bf 2017}, {\em 843},~62.
 [\href{http://dx.doi.org/10.3847/1538-4357/aa72f5}{CrossRef}]

\bibitem[{Shao} {et~al.}(2019){Shao}, {Cautun}, and {Frenk}]{Shao2019}
{Shao}, S.; {Cautun}, M.; {Frenk}, C.S.
\newblock {Evolution of galactic planes of satellites in the EAGLE simulation}.
\newblock {\em Mon. Not. R. Astron. Soc.} {\bf 2019}, {\em 488},~1166--1179.
[\href{http://dx.doi.org/10.1093/mnras/stz1741}{CrossRef}]

\bibitem[{Gu} {et~al.}(2022){Gu}, {Guo}, {Zhang}, {Cautun}, {Lacey},
{Frenk}, and {Shao}]{Gu2022}
{Gu}, Q.; {Guo}, Q.; {Zhang}, T.; {Cautun}, M.; {Lacey}, C.; {Frenk}, C.S.;
{Shao}, S.
\newblock {The spatial distribution of satellites in galaxy clusters}.
\newblock {\em Mon. Not. R. Astron. Soc.} {\bf 2022}, {\em 514},~390--402.
 [\href{http://dx.doi.org/10.1093/mnras/stac1292}{CrossRef}]

\bibitem[{Pawlowski} {et~al.}(2012){Pawlowski}, {Kroupa}, {Angus}, {de
Boer}, {Famaey}, and {Hensler}]{Pawlowski:2012a}
{Pawlowski}, M.S.; {Kroupa}, P.; {Angus}, G.; {de Boer}, K.S.; {Famaey}, B.;
{Hensler}, G.
\newblock {Filamentary accretion cannot explain the orbital poles of the Milky
Way satellites}.
\newblock {\em Mon. Not. R. Astron. Soc.} {\bf 2012}, {\em 424},~80--92.
[\href{http://dx.doi.org/10.1111/j.1365-2966.2012.21169.x}{CrossRef}]

\bibitem[{Cautun} and {Frenk}(2017)]{Cautun2017}
{Cautun}, M.; {Frenk}, C.S.
\newblock {The tangential velocity excess of the Milky Way satellites}.
\newblock {\em Mon. Not. R. Astron. Soc.} {\bf 2017}, {\em 468},~L41--L45.
[\href{http://dx.doi.org/10.1093/mnrasl/slx025}{CrossRef}]

\bibitem[{Sawala} {et~al.}(2022){Sawala}, {Cautun}, {Frenk}, {Helly},
{Jasche}, {Jenkins}, {Johansson}, {Lavaux}, {McAlpine}, and
{Schaller}]{Sawala2022}
{Sawala}, T.; {Cautun}, M.; {Frenk}, C.; {Helly}, J.; {Jasche}, J.; {Jenkins},
A.; {Johansson}, P.H.; {Lavaux}, G.; {McAlpine}, S.; {Schaller}, M.
\newblock {The Milky Way's plane of satellites is consistent with
{\ensuremath{\Lambda}}CDM}.
\newblock {\em Nat. Astron.} {\bf 2022}, {\emph{7}, 481--491}.
 [\href{http://dx.doi.org/10.1038/s41550-022-01856-z}{CrossRef}]

\bibitem[{Holmberg}(1969)]{Holmberg:1969}
{Holmberg}, E.
\newblock {A study of physical groups of galaxies}.
\newblock {\em Ark. Astron.} {\bf 1969}, {\em 5},~305--343.

\bibitem[{Zaritsky} {et~al.}(1997){Zaritsky}, {Smith}, {Frenk}, and
{White}]{Zaritsky:1997}
{Zaritsky}, D.; {Smith}, R.; {Frenk}, C.S.; {White}, S.D.M.
\newblock {Anisotropies in the Distribution of Satellite Galaxies}.
\newblock {\em Astrophys. J. Lett.} {\bf 1997}, {\em 478},~L53.
 [\href{http://dx.doi.org/10.1086/310557}{CrossRef}]

\bibitem[{Sales} and {Lambas}(2004)]{Sales:2004}
{Sales}, L.; {Lambas}, D.G.
\newblock {Anisotropy in the distribution of satellites around primary galaxies
in the 2dF Galaxy Redshift Survey: The Holmberg effect}.
\newblock {\em Mon. Not. R. Astron. Soc.} {\bf 2004}, {\em 348},~1236--1240.
 [\href{http://dx.doi.org/10.1111/j.1365-2966.2004.07443.x}{CrossRef}]

\bibitem[{Brainerd}(2005)]{Brainerd:2005}
{Brainerd}, T.G.
\newblock {Anisotropic Distribution of SDSS Satellite Galaxies: Planar (Not
Polar) Alignment}.
\newblock {\em Astrophys. J. Lett.} {\bf 2005}, {\em 628},~L101--L104.
 [\href{http://dx.doi.org/10.1086/432713}{CrossRef}]

\bibitem[{Yang} {et~al.}(2005){Yang}, {Mo}, {van den Bosch}, {Weinmann},
{Li}, and {Jing}]{Yang:2005}
{Yang}, X.; {Mo}, H.J.; {van den Bosch}, F.C.; {Weinmann}, S.M.; {Li}, C.;
{Jing}, Y.P.
\newblock {The cross-correlation between galaxies and groups: Probing the
galaxy distribution in and around dark matter haloes}.
\newblock {\em Mon. Not. R. Astron. Soc.} {\bf 2005}, {\em 362},~711--726.
 [\href{http://dx.doi.org/10.1111/j.1365-2966.2005.09351.x}{CrossRef}]

\bibitem[{Azzaro} {et~al.}(2007){Azzaro}, {Patiri}, {Prada}, and
{Zentner}]{Azzaro:2007}
{Azzaro}, M.; {Patiri}, S.G.; {Prada}, F.; {Zentner}, A.R.
\newblock {Angular distribution of satellite galaxies from the Sloan Digital
Sky Survey Data Release 4}.
\newblock {\em Mon. Not. R. Astron. Soc.} {\bf 2007}, {\em 376},~L43--L47.
 [\href{http://dx.doi.org/10.1111/j.1745-3933.2007.00282.x}{CrossRef}]

\bibitem[{Bailin} {et~al.}(2008){Bailin}, {Power}, {Norberg}, {Zaritsky},
and {Gibson}]{Bailin:2008}
{Bailin}, J.; {Power}, C.; {Norberg}, P.; {Zaritsky}, D.; {Gibson}, B.K.
\newblock {The anisotropic distribution of satellite galaxies}.
\newblock {\em Mon. Not. R. Astron. Soc.} {\bf 2008}, {\em 390},~1133--1156.
 [\href{http://dx.doi.org/10.1111/j.1365-2966.2008.13828.x}{CrossRef}]

\bibitem[{Steffen} and {Valenzuela}(2008)]{Steffen:2008}
{Steffen}, J.H.; {Valenzuela}, O.
\newblock {Constraints on the angular distribution of satellite galaxies about
spiral hosts}.
\newblock {\em Mon. Not. R. Astron. Soc.} {\bf 2008}, {\em 387},~1199--1205.
 [\href{http://dx.doi.org/10.1111/j.1365-2966.2008.13314.x}{CrossRef}]

\bibitem[{Agustsson} and {Brainerd}(2010)]{Agustsson:2010}
{Agustsson}, I.; {Brainerd}, T.G.
\newblock {Anisotropic Locations of Satellite Galaxies: Clues to the
Orientations of Galaxies within their Dark Matter Halos}.
\newblock {\em Astrophys. J.} {\bf 2010}, {\em 709},~1321--1336.
 [\href{http://dx.doi.org/10.1088/0004-637X/709/2/1321}{CrossRef}]

\bibitem[{Kang} {et~al.}(2005){Kang}, {Mao}, {Gao}, and {Jing}]{Kang:2005}
{Kang}, X.; {Mao}, S.; {Gao}, L.; {Jing}, Y.P.
\newblock {Are great disks defined by satellite galaxies in Milky-Way type
halos rare in {$\Lambda$}CDM?}
\newblock {\em Astron. Astrophys.} {\bf 2005}, {\em 437},~383--388.
 [\href{http://dx.doi.org/10.1051/0004-6361:20052675}{CrossRef}]

\bibitem[{Libeskind} {et~al.}(2005){Libeskind}, {Frenk}, {Cole}, {Helly},
{Jenkins}, {Navarro}, and {Power}]{Libeskind:2005}
{Libeskind}, N.I.; {Frenk}, C.S.; {Cole}, S.; {Helly}, J.C.; {Jenkins}, A.;
{Navarro}, J.F.; {Power}, C.
\newblock {The distribution of satellite galaxies: The great pancake}.
\newblock {\em Mon. Not. R. Astron. Soc.} {\bf 2005}, {\em 363},~146--152.
[\href{http://dx.doi.org/10.1111/j.1365-2966.2005.09425.x}{CrossRef}]

\bibitem[{Zentner} {et~al.}(2005){Zentner}, {Kravtsov}, {Gnedin}, and
{Klypin}]{Zentner:2005}
{Zentner}, A.R.; {Kravtsov}, A.V.; {Gnedin}, O.Y.; {Klypin}, A.A.
\newblock {The Anisotropic Distribution of Galactic Satellites}.
\newblock {\em Astrophys. J.} {\bf 2005}, {\em 629},~219--232.
[\href{http://dx.doi.org/10.1086/431355}{CrossRef}]

\bibitem[{Agustsson} and {Brainerd}(2006)]{Agustsson:2006}
{Agustsson}, I.; {Brainerd}, T.G.
\newblock {The Locations of Satellite Galaxies in a {$\Lambda$}CDM Universe}.
\newblock {\em Astrophys. J.} {\bf 2006}, {\em 650},~550--559.
 [\href{http://dx.doi.org/10.1086/507084}{CrossRef}]

\bibitem[{Libeskind} {et~al.}(2007){Libeskind}, {Cole}, {Frenk}, {Okamoto},
and {Jenkins}]{Libeskind:2007}
{Libeskind}, N.I.; {Cole}, S.; {Frenk}, C.S.; {Okamoto}, T.; {Jenkins}, A.
\newblock {Satellite systems around galaxies in hydrodynamic simulations}.
\newblock {\em Mon. Not. R. Astron. Soc.} {\bf 2007}, {\em 374},~16--28.
 [\href{http://dx.doi.org/10.1111/j.1365-2966.2006.11205.x}{CrossRef}]

\bibitem[{Libeskind} {et~al.}(2009){Libeskind}, {Frenk}, {Cole}, {Jenkins},
and {Helly}]{Libeskind:2009}
{Libeskind}, N.I.; {Frenk}, C.S.; {Cole}, S.; {Jenkins}, A.; {Helly}, J.C.
\newblock {How common is the Milky Way-satellite system alignment?}
\newblock {\em Mon. Not. R. Astron. Soc.} {\bf 2009}, {\em 399},~550--558.
 [\href{http://dx.doi.org/10.1111/j.1365-2966.2009.15315.x}{CrossRef}]

\bibitem[{Deason} {et~al.}(2011){Deason}, {McCarthy}, {Font}, {Evans},
{Frenk}, {Belokurov}, {Libeskind}, {Crain}, and {Theuns}]{Deason:2011}
{Deason}, A.J.; {McCarthy}, I.G.; {Font}, A.S.; {Evans}, N.W.; {Frenk}, C.S.;
{Belokurov}, V.; {Libeskind}, N.I.; {Crain}, R.A.; {Theuns}, T.
\newblock {Mismatch and misalignment: Dark haloes and satellites of disc
galaxies}.
\newblock {\em Mon. Not. R. Astron. Soc.} {\bf 2011}, {\em 415},~2607--2625.
 [\href{http://dx.doi.org/10.1111/j.1365-2966.2011.18884.x}{CrossRef}]

\bibitem[{Wang} {et~al.}(2013){Wang}, {Frenk}, and {Cooper}]{Wang:2012}
{Wang}, J.; {Frenk}, C.S.; {Cooper}, A.P.
\newblock {The spatial distribution of galactic satellites in the
{\ensuremath{\Lambda}} cold dark matter cosmology}.
\newblock {\em Mon. Not. R. Astron. Soc.} {\bf 2013}, {\em 429},~1502--1513.
 [\href{http://dx.doi.org/10.1093/mnras/sts442}{CrossRef}]

\bibitem[{Lovell} {et~al.}(2011){Lovell}, {Eke}, {Frenk}, and
{Jenkins}]{Lovell:2011}
{Lovell}, M.R.; {Eke}, V.R.; {Frenk}, C.S.; {Jenkins}, A.
\newblock {The link between galactic satellite orbits and sub-halo accretion}.
\newblock {\em Mon. Not. R. Astron. Soc.} {\bf 2011}, {\em 413},~3013--3021.
 [\href{http://dx.doi.org/10.1111/j.1365-2966.2011.18377.x}{CrossRef}]

\bibitem[{Pawlowski} and {Kroupa}(2020)]{Pawlowski:2020}
{Pawlowski}, M.S.; {Kroupa}, P.
\newblock {The Milky Way's disc of classical satellite galaxies in light of
Gaia DR2}.
\newblock {\em Mon. Not. R. Astron. Soc.} {\bf 2020}, {\em 491},~3042--3059.
 [\href{http://dx.doi.org/10.1093/mnras/stz3163}{CrossRef}]

\bibitem[{Boylan-Kolchin}(2021)]{Boylan:2021}
{Boylan-Kolchin}, M.
\newblock {Planes of satellites are not a problem for (just)
{\ensuremath{\Lambda}}CDM}.
\newblock {\em Nat. Astron.} {\bf 2021}, {\em 5},~1188--1190.
 [\href{http://dx.doi.org/10.1038/s41550-021-01467-0}{CrossRef}]

\bibitem[{Zwicky}(1956)]{Zwicky:1956}
{Zwicky}, F.
\newblock {Multiple Galaxies}.
\newblock {\em Ergeb. Exakten Naturwiss.} {\bf 1956}, {\em
29},~344--385.

\bibitem[{Kroupa}(1997)]{Kroupa:1997}
{Kroupa}, P.
\newblock {Dwarf spheroidal satellite galaxies without dark matter}.
\newblock {\em New Astron.} {\bf 1997}, {\em 2},~139--164.
 [\href{http://dx.doi.org/10.1016/S1384-1076(97)00012-2}{CrossRef}]

\bibitem[{Metz} and {Kroupa}(2007)]{Metz:2007a}
{Metz}, M.; {Kroupa}, P.
\newblock {Dwarf spheroidal satellites: Are they of tidal origin?}
\newblock {\em Mon. Not. R. Astron. Soc.} {\bf 2007}, {\em 376},~387--392.
 [\href{http://dx.doi.org/10.1111/j.1365-2966.2007.11438.x}{CrossRef}]

\bibitem[{Okazaki} and {Taniguchi}(2000)]{Okazaki:2000}
{Okazaki}, T.; {Taniguchi}, Y.
\newblock {Dwarf Galaxy Formation Induced by Galaxy Interactions}.
\newblock {\em Astrophys. J.} {\bf 2000}, {\em 543},~149--152.
 [\href{http://dx.doi.org/10.1086/317109}{CrossRef}]

\bibitem[{Bournaud}(2010)]{Bournaud:2010}
{Bournaud}, F.
\newblock {Tidal Dwarf Galaxies and Missing Baryons}.
\newblock {\em Adv. Astron.} {\bf 2010}, {\em 2010}, {735284}.
 [\href{http://dx.doi.org/10.1155/2010/735284}{CrossRef}]

\bibitem[{Pawlowski} {et~al.}(2011){Pawlowski}, {Kroupa}, and {de
Boer}]{Pawlowski:2011}
{Pawlowski}, M.S.; {Kroupa}, P.; {de Boer}, K.S.
\newblock {Making counter-orbiting tidal debris. The origin of the Milky Way
disc of satellites?}
\newblock {\em Astron. Astrophys.} {\bf 2011}, {\em 532},~A118.
[\href{http://dx.doi.org/10.1051/0004-6361/201015021}{CrossRef}]

\bibitem[{Wetzstein} {et~al.}(2007){Wetzstein}, {Naab}, and
{Burkert}]{Wetzstein:2007}
{Wetzstein}, M.; {Naab}, T.; {Burkert}, A.
\newblock {Do dwarf galaxies form in tidal tails?}
\newblock {\em Mon. Not. R. Astron. Soc.} {\bf 2007}, {\em 375},~805--820.
 [\href{http://dx.doi.org/10.1111/j.1365-2966.2006.11360.x}{CrossRef}]

\bibitem[{Hammer} {et~al.}(2010){Hammer}, {Yang}, {Wang}, {Puech}, {Flores},
and {Fouquet}]{Hammer:2010}
{Hammer}, F.; {Yang}, Y.B.; {Wang}, J.L.; {Puech}, M.; {Flores}, H.; {Fouquet},
S.
\newblock {Does M31 Result from an Ancient Major Merger?}
\newblock {\em Astrophys. J.} {\bf 2010}, {\em 725},~542--555.
 [\href{http://dx.doi.org/10.1088/0004-637X/725/1/542}{CrossRef}]

\bibitem[{Yang} and {Hammer}(2010)]{Yang:2010}
{Yang}, Y.; {Hammer}, F.
\newblock {Could the Magellanic Clouds be Tidal Dwarfs Expelled from a
Past-merger Event Occurring in Andromeda?}
\newblock {\em Astrophys. J. Lett.} {\bf 2010}, {\em 725},~L24--L27.
 [\href{http://dx.doi.org/10.1088/2041-8205/725/1/L24}{CrossRef}]

\bibitem[{Fouquet} {et~al.}(2012){Fouquet}, {Hammer}, {Yang}, {Puech}, and
{Flores}]{Fouquet:2012}
{Fouquet}, S.; {Hammer}, F.; {Yang}, Y.; {Puech}, M.; {Flores}, H.
\newblock {Does the dwarf galaxy system of the Milky Way originate from
Andromeda?}
\newblock {\em {Mon. Not. R. Astron. Soc.}} {\bf 2012}, {\emph{427}, 1769--1783}.
 [\href{http://dx.doi.org/10.1111/j.1365-2966.2012.22067.x}{CrossRef}]

\bibitem[{Barnes} and {Hernquist}(1992)]{Barnes:1992}
{Barnes}, J.E.; {Hernquist}, L.
\newblock {Formation of dwarf galaxies in tidal tails}.
\newblock {\em Nature} {\bf 1992}, {\em 360},~715--717.
 [\href{http://dx.doi.org/10.1038/360715a0}{CrossRef}]

\bibitem[{Simon} and {Geha}(2007)]{Simon:2007}
{Simon}, J.D.; {Geha}, M.
\newblock {The Kinematics of the Ultra-faint Milky Way Satellites: Solving the
Missing Satellite Problem}.
\newblock {\em Astrophys. J.} {\bf 2007}, {\em 670},~313--331.
 [\href{http://dx.doi.org/10.1086/521816}{CrossRef}]

\bibitem[{Simon} {et~al.}(2011){Simon}, {Geha}, {Minor}, {Martinez},
{Kirby}, {Bullock}, {Kaplinghat}, {Strigari}, {Willman}, {Choi}, {Tollerud},
and {Wolf}]{Simon:2011}
{Simon}, J.D.; {Geha}, M.; {Minor}, Q.E.; {Martinez}, G.D.; {Kirby}, E.N.;
{Bullock}, J.S.; {Kaplinghat}, M.; {Strigari}, L.E.; {Willman}, B.; {Choi},
P.I.;  et~al.
\newblock {A Complete Spectroscopic Survey of the Milky Way Satellite Segue 1:
The Darkest Galaxy}.
\newblock {\em Astrophys. J.} {\bf 2011}, {\em 733},~46.
[\href{http://dx.doi.org/10.1088/0004-637X/733/1/46}{CrossRef}]

\bibitem[{Klessen} and {Kroupa}(1998)]{Klessen:1998}
{Klessen}, R.S.; {Kroupa}, P.
\newblock {Dwarf Spheroidal Satellite Galaxies without Dark Matter: Results
from Two Different Numerical Techniques}.
\newblock {\em Astrophys. J.} {\bf 1998}, {\em 498},~143.
 [\href{http://dx.doi.org/10.1086/305540}{CrossRef}]

\bibitem[{Casas} {et~al.}(2012){Casas}, {Arias}, {Pe{\~n}a Ram{\'{\i}}rez},
and {Kroupa}]{Casas:2012}
{Casas}, R.A.; {Arias}, V.; {Pe{\~n}a Ram{\'{\i}}rez}, K.; {Kroupa}, P.
\newblock {Dwarf spheroidal satellites of the Milky Way from dark matter free
tidal dwarf galaxy progenitors: Maps of orbits}.
\newblock {\em Mon. Not. R. Astron. Soc.} {\bf 2012}, {\em 424},~1941--1951.
 [\href{http://dx.doi.org/10.1111/j.1365-2966.2012.21319.x}{CrossRef}]

\bibitem[{Samuel} {et~al.}(2021){Samuel}, {Wetzel}, {Chapman}, {Tollerud},
{Hopkins}, {Boylan-Kolchin}, {Bailin}, and {Faucher-Gigu{\`e}re}]{Samuel21}
{Samuel}, J.; {Wetzel}, A.; {Chapman}, S.; {Tollerud}, E.; {Hopkins}, P.F.;
{Boylan-Kolchin}, M.; {Bailin}, J.; {Faucher-Gigu{\`e}re}, C.A.
\newblock {Planes of satellites around Milky Way/M31-mass galaxies in the FIRE
simulations and comparisons with the Local Group}.
\newblock {\em Mon. Not. R. Astron. Soc.} {\bf 2021}, {\em 504},~1379--1397.
 [\href{http://dx.doi.org/10.1093/mnras/stab955}{CrossRef}]

\bibitem[{Wetzel} {et~al.}(2023){Wetzel}, {Hayward}, {Sanderson}, {Ma},
{Angl{\'e}s-Alc{\'a}zar}, {Feldmann}, {Chan}, {El-Badry}, {Wheeler},
{Garrison-Kimmel}, {Nikakhtar}, {Panithanpaisal}, {Arora}, {Gurvich},
{Samuel}, {Sameie}, {Pandya}, {Hafen}, {Hummels}, {Loebman},
{Boylan-Kolchin}, {Bullock}, {Faucher-Gigu{\`e}re}, {Kere{\v{s}}},
{Quataert}, and {Hopkins}]{Wetzel23}
{Wetzel}, A.; {Hayward}, C.C.; {Sanderson}, R.E.; {Ma}, X.;
{Angl{\'e}s-Alc{\'a}zar}, D.; {Feldmann}, R.; {Chan}, T.K.; {El-Badry}, K.;
{Wheeler}, C.; {Garrison-Kimmel}, S.;  et~al.
\newblock {Public Data Release of the FIRE-2 Cosmological Zoom-in Simulations
of Galaxy Formation}.
\newblock {\em Astrophys. J. Suppl. Ser.} {\bf 2023}, {\em 265},~44.
 [\href{http://dx.doi.org/10.3847/1538-4365/acb99a}{CrossRef}]

\bibitem[{Vasiliev}(2023{\natexlab{a}})]{Vasiliev:2023a}
{Vasiliev}, E.
\newblock {The Effect of the LMC on the Milky Way System}.
\newblock {\em Galaxies} {\bf 2023}, {\em 11},~59.
 [\href{http://dx.doi.org/10.3390/galaxies11020059}{CrossRef}]

\bibitem[{Vasiliev}(2023{\natexlab{b}})]{Vasiliev:2023b}
{Vasiliev}, E.
\newblock {Dear Magellanic Clouds, welcome back!}
\newblock {\em arXiv} {\bf 2023}, arXiv:2306.04837.
 [\href{https://doi.org/10.48550/arXiv.2306.04837}{CrossRef}]

\bibitem[Förster {et~al.}(2022)Förster, Remus, Dolag, Kimmig, Teklu, and
Valenzuela]{Forster22}
Förster, P.U.; Remus, R.S.; Dolag, K.; Kimmig, L.C.; Teklu, A.; Valenzuela,
L.M.
\newblock Planes of Satellite Galaxies in the Magneticum Pathfinder
Simulations. \emph{arXiv} \textbf{2022},  arXiv:2208.05496.
 [\href{https://doi.org/10.48550/arXiv.2208.05496}{CrossRef}]

\bibitem[{Libeskind} {et~al.}(2020){Libeskind}, {Carlesi}, {Grand},
{Khalatyan}, {Knebe}, {Pakmor}, {Pilipenko}, {Pawlowski}, {Sparre}, {Tempel},
{Wang}, {Courtois}, {Gottl{\"o}ber}, {Hoffman}, {Minchev}, {Pfrommer},
{Sorce}, {Springel}, {Steinmetz}, {Tully}, {Vogelsberger}, and
{Yepes}]{Libeskind20}
{Libeskind}, N.I.; {Carlesi}, E.; {Grand}, R.J.J.; {Khalatyan}, A.; {Knebe},
A.; {Pakmor}, R.; {Pilipenko}, S.; {Pawlowski}, M.S.; {Sparre}, M.; {Tempel},
E.;  et~al.
\newblock {The HESTIA project: Simulations of the Local Group}.
\newblock {\em Mon. Not. R. Astron. Soc.} {\bf 2020}, {\em 498},~2968--2983.
 [\href{http://dx.doi.org/10.1093/mnras/staa2541}{CrossRef}]

\bibitem[{Dupuy} {et~al.}(2022){Dupuy}, {Libeskind}, {Hoffman}, {Courtois},
{Gottl{\"o}ber}, {Grand}, {Knebe}, {Sorce}, {Tempel}, {Tully},
{Vogelsberger}, and {Wang}]{Dupuy22}
{Dupuy}, A.; {Libeskind}, N.I.; {Hoffman}, Y.; {Courtois}, H.M.;
{Gottl{\"o}ber}, S.; {Grand}, R.J.J.; {Knebe}, A.; {Sorce}, J.G.; {Tempel},
E.; {Tully}, R.B.;  et~al.
\newblock {Anisotropic satellite accretion on to the Local Group with HESTIA}.
\newblock {\em Mon. Not. R. Astron. Soc.} {\bf 2022}, {\em 516},~4576--4584.
 [\href{http://dx.doi.org/10.1093/mnras/stac2486}{CrossRef}]

\bibitem[{Pham} {et~al.}(2023){Pham}, {Kravtsov}, and {Manwadkar}]{Pham23}
{Pham}, K.; {Kravtsov}, A.; {Manwadkar}, V.
\newblock {Spatial and orbital planes of the Milky Way satellites: Unusual but
consistent with {\ensuremath{\Lambda}}CDM}.
\newblock {\em Mon. Not. R. Astron. Soc.} {\bf 2023}, {\em 520},~3937--3946.
 [\href{http://dx.doi.org/10.1093/mnras/stad335}{CrossRef}]

\bibitem[Xu {et~al.}(2023)Xu, Kang, and Libeskind]{Xu23}
Xu, Y.; Kang, X.; Libeskind, N.I.
\newblock {A rotating satellite plane around Milky Way-like galaxy from the TNG50 simulation}.  \emph{arXiv} \textbf{2023}, arXiv:2303.00441.
 [\href{https://doi.org/10.48550/arXiv.2303.00441}{CrossRef}]

\bibitem[{Scannapieco} {et~al.}(2009){Scannapieco}, {White}, {Springel}, and
{Tissera}]{Scannapieco:2009}
{Scannapieco}, C.; {White}, S.D.M.; {Springel}, V.; {Tissera}, P.B.
\newblock {The formation and survival of discs in a {$\Lambda$}CDM universe}.
\newblock {\em Mon. Not. R. Astron. Soc.} {\bf 2009}, {\em 396},~696--708.
 [\href{http://dx.doi.org/10.1111/j.1365-2966.2009.14764.x}{CrossRef}]

\bibitem[{Scannapieco} {et~al.}(2011){Scannapieco}, {White}, {Springel}, and
{Tissera}]{Scannapieco:2011}
{Scannapieco}, C.; {White}, S.D.M.; {Springel}, V.; {Tissera}, P.B.
\newblock {Formation history, structure and dynamics of discs and spheroids in
simulated Milky Way mass galaxies}.
\newblock {\em Mon. Not. R. Astron. Soc.} {\bf 2011}, \emph{417}, 154--171.
 [\href{http://dx.doi.org/10.1111/j.1365-2966.2011.19027.x}{CrossRef}]

\bibitem[{Springel} {et~al.}(2008){Springel}, {Wang}, {Vogelsberger},
{Ludlow}, {Jenkins}, {Helmi}, {Navarro}, {Frenk}, and {White}]{Springel:2008}
{Springel}, V.; {Wang}, J.; {Vogelsberger}, M.; {Ludlow}, A.; {Jenkins}, A.;
{Helmi}, A.; {Navarro}, J.F.; {Frenk}, C.S.; {White}, S.D.M.
\newblock {The Aquarius Project: The sub-haloes of galactic haloes}.
\newblock {\em Mon. Not. R. Astron. Soc.} {\bf 2008}, {\em 391},~1685--1711.
[\href{http://dx.doi.org/10.1111/j.1365-2966.2008.14066.x}{CrossRef}]

\bibitem[{Springel}(2005)]{Springel:2005}
{Springel}, V.
\newblock {The cosmological simulation code GADGET-2}.
\newblock {\em Mon. Not. R. Astron. Soc.} {\bf 2005}, {\em 364},~1105--1134.
 [\href{http://dx.doi.org/10.1111/j.1365-2966.2005.09655.x}{CrossRef}]

\bibitem[{Marinacci} {et~al.}(2014){Marinacci}, {Pakmor}, and
{Springel}]{Marinacci14}
{Marinacci}, F.; {Pakmor}, R.; {Springel}, V.
\newblock {The formation of disc galaxies in high-resolution moving-mesh
cosmological simulations}.
\newblock {\em Mon. Not. R. Astron. Soc.} {\bf 2014}, {\em 437},~1750--1775.
 [\href{http://dx.doi.org/10.1093/mnras/stt2003}{CrossRef}]

\bibitem[{Sawala} {et~al.}(2012){Sawala}, {Scannapieco}, and
{White}]{Sawala12}
{Sawala}, T.; {Scannapieco}, C.; {White}, S.
\newblock {Local Group dwarf galaxies: Nature and nurture}.
\newblock {\em Mon. Not. R. Astron. Soc.} {\bf 2012}, {\em 420},~1714--1730.
 [\href{http://dx.doi.org/10.1111/j.1365-2966.2011.20181.x}{CrossRef}]

\bibitem[{Springel} {et~al.}(2001){Springel}, {White}, {Tormen}, and
{Kauffmann}]{Springel:2001a}
{Springel}, V.; {White}, S.D.M.; {Tormen}, G.; {Kauffmann}, G.
\newblock {Populating a cluster of galaxies---I. Results at \emph{z} = 0}.
\newblock {\em Mon. Not. R. Astron. Soc.} {\bf 2001}, {\em 328},~726--750.
 [\href{http://dx.doi.org/10.1046/j.1365-8711.2001.04912.x}{CrossRef}]

\bibitem[{Klypin} {et~al.}(1999){Klypin}, {Kravtsov}, {Valenzuela}, and
{Prada}]{Klypin:1999}
{Klypin}, A.; {Kravtsov}, A.V.; {Valenzuela}, O.; {Prada}, F.
\newblock {Where Are the Missing Galactic Satellites?}
\newblock {\em Astrophys. J.} {\bf 1999}, {\em 522},~82--92.
 [\href{http://dx.doi.org/10.1086/307643}{CrossRef}]

\bibitem[{Moore} {et~al.}(1999){Moore}, {Ghigna}, {Governato}, {Lake},
{Quinn}, {Stadel}, and {Tozzi}]{Moore:1999}
{Moore}, B.; {Ghigna}, S.; {Governato}, F.; {Lake}, G.; {Quinn}, T.; {Stadel},
J.; {Tozzi}, P.
\newblock {Dark Matter Substructure within Galactic Halos}.
\newblock {\em Astrophys. J. Lett.} {\bf 1999}, {\em 524},~L19--L22.
 [\href{http://dx.doi.org/10.1086/312287}{CrossRef}]

\bibitem[{Wadepuhl} and {Springel}(2011)]{Wadepuhl:2011}
{Wadepuhl}, M.; {Springel}, V.
\newblock {Satellite galaxies in hydrodynamical simulations of Milky Way sized
galaxies}.
\newblock {\em Mon. Not. R. Astron. Soc.} {\bf 2011}, {\em 410},~1975--1992.
 [\href{http://dx.doi.org/10.1111/j.1365-2966.2010.17576.x}{CrossRef}]

\bibitem[{Font} {et~al.}(2011){Font}, {Benson}, {Bower}, {Frenk}, {Cooper},
{De Lucia}, {Helly}, {Helmi}, {Li}, {McCarthy}, {Navarro}, {Springel},
{Starkenburg}, {Wang}, and {White}]{Font:2011}
{Font}, A.S.; {Benson}, A.J.; {Bower}, R.G.; {Frenk}, C.S.; {Cooper}, A.; {De
Lucia}, G.; {Helly}, J.C.; {Helmi}, A.; {Li}, Y.S.; {McCarthy}, I.G.;  et~al.
\newblock {The population of Milky Way satellites in the {$\Lambda$} cold dark
matter cosmology}.
\newblock {\em Mon. Not. R. Astron. Soc.} {\bf 2011}, {\em 417},~1260--1279.
 [\href{http://dx.doi.org/10.1111/j.1365-2966.2011.19339.x}{CrossRef}]

\bibitem[{Santos-Santos} {et~al.}(2022){Santos-Santos}, {Sales}, {Fattahi},
and {Navarro}]{Santos22}
{Santos-Santos}, I.M.E.; {Sales}, L.V.; {Fattahi}, A.; {Navarro}, J.F.
\newblock {Satellite mass functions and the faint end of the galaxy mass-halo
mass relation in LCDM}.
\newblock {\em Mon. Not. R. Astron. Soc.} {\bf 2022}, {\em 515},~3685--3697.
 [\href{http://dx.doi.org/10.1093/mnras/stac2057}{CrossRef}]

\bibitem[{Gaia Collaboration} {et~al.}(2018){Gaia Collaboration}, {Helmi},
{van Leeuwen}, {McMillan}, {Massari}, {Antoja}, {Robin}, {Lindegren},
{Bastian}, {Arenou}, {Babusiaux}, {Biermann}, {Breddels}, {Hobbs}, {Jordi},
{Pancino}, {Reyl{\'e}}, {Veljanoski}, {Brown}, {Vallenari}, {Prusti}, {de
Bruijne}, {Bailer-Jones}, {Evans}, {Eyer}, {Jansen}, {Klioner}, {Lammers},
{Luri}, {Mignard}, {Panem}, {Pourbaix}, {Randich}, {Sartoretti}, {Siddiqui},
{Soubiran}, {Walton}, {Cropper}, {Drimmel}, {Katz}, {Lattanzi}, {Bakker},
{Cacciari}, {Casta{\~n}eda}, {Chaoul}, {Cheek}, {De Angeli}, {Fabricius},
{Guerra}, {Holl}, {Masana}, {Messineo}, {Mowlavi}, {Nienartowicz}, {Panuzzo},
{Portell}, {Riello}, {Seabroke}, {Tanga}, {Th{\'e}venin}, {Gracia-Abril},
{Comoretto}, {Garcia-Reinaldos}, {Teyssier}, {Altmann}, {Andrae}, {Audard},
{Bellas-Velidis}, {Benson}, {Berthier}, {Blomme}, {Burgess}, {Busso},
{Carry}, {Cellino}, {Clementini}, {Clotet}, {Creevey}, {Davidson}, {De
Ridder}, {Delchambre}, {Dell'Oro}, {Ducourant},
{Fern{\'a}ndez-Hern{\'a}ndez}, {Fouesneau}, {Fr{\'e}mat}, {Galluccio},
{Garc{\'\i}a-Torres}, {Gonz{\'a}lez-N{\'u}{\~n}ez}, {Gonz{\'a}lez-Vidal},
{Gosset}, {Guy}, {Halbwachs}, {Hambly}, {Harrison}, {Hern{\'a}ndez},
{Hestroffer}, {Hodgkin}, {Hutton}, {Jasniewicz}, {Jean-Antoine-Piccolo},
{Jordan}, {Korn}, {Krone-Martins}, {Lanzafame}, {Lebzelter}, {L{\"o}ffler},
{Manteiga}, {Marrese}, {Mart{\'\i}n-Fleitas}, {Moitinho}, {Mora}, {Muinonen},
{Osinde}, {Pauwels}, {Petit}, {Recio-Blanco}, {Richards}, {Rimoldini},
{Sarro}, {Siopis}, {Smith}, {Sozzetti}, {S{\"u}veges}, {Torra}, {van Reeven},
{Abbas}, {Abreu Aramburu}, {Accart}, {Aerts}, {Altavilla}, {{\'A}lvarez},
{Alvarez}, {Alves}, {Anderson}, {Andrei}, {Anglada Varela}, {Antiche},
{Arcay}, {Astraatmadja}, {Bach}, {Baker}, {Balaguer-N{\'u}{\~n}ez}, {Balm},
{Barache}, {Barata}, {Barbato}, {Barblan}, {Barklem}, {Barrado}, {Barros},
{Barstow}, {Bartholom{\'e} Mu{\~n}oz}, {Bassilana}, {Becciani}, {Bellazzini},
{Berihuete}, {Bertone}, {Bianchi}, {Bienaym{\'e}}, {Blanco-Cuaresma}, {Boch},
{Boeche}, {Bombrun}, {Borrachero}, {Bossini}, {Bouquillon}, {Bourda},
{Bragaglia}, {Bramante}, {Bressan}, {Brouillet}, {Br{\"u}semeister},
{Brugaletta}, {Bucciarelli}, {Burlacu}, {Busonero}, {Butkevich}, {Buzzi},
{Caffau}, {Cancelliere}, {Cannizzaro}, {Cantat-Gaudin}, {Carballo},
{Carlucci}, {Carrasco}, {Casamiquela}, {Castellani}, {Castro-Ginard},
{Charlot}, {Chemin}, {Chiavassa}, {Cocozza}, {Costigan}, {Cowell}, {Crifo},
{Crosta}, {Crowley}, {Cuypers}, {Dafonte}, {Damerdji}, {Dapergolas}, {David},
{David}, {de Laverny}, {De Luise}, {De March}, {de Martino}, {de Souza}, {de
Torres}, {Debosscher}, {del Pozo}, {Delbo}, {Delgado}, {Delgado}, {Di
Matteo}, {Diakite}, {Diener}, {Distefano}, {Dolding}, {Drazinos},
{Dur{\'a}n}, {Edvardsson}, {Enke}, {Eriksson}, {Esquej}, {Eynard Bontemps},
{Fabre}, {Fabrizio}, {Faigler}, {Falc{\~a}o}, {Farr{\`a}s Casas}, {Federici},
{Fedorets}, {Fernique}, {Figueras}, {Filippi}, {Findeisen}, {Fonti},
{Fraile}, {Fraser}, {Fr{\'e}zouls}, {Gai}, {Galleti}, {Garabato},
{Garc{\'\i}a-Sedano}, {Garofalo}, {Garralda}, {Gavel}, {Gavras}, {Gerssen},
{Geyer}, {Giacobbe}, {Gilmore}, {Girona}, {Giuffrida}, {Glass}, {Gomes},
{Granvik}, {Gueguen}, {Guerrier}, {Guiraud}, {Guti{\'e}rrez-S{\'a}nchez},
{Hofmann}, {Holland}, {Huckle}, {Hypki}, {Icardi}, {Jan{\ss}en}, {Jevardat de
Fombelle}, {Jonker}, {Juh{\'a}sz}, {Julbe}, {Karampelas}, {Kewley}, {Klar},
{Kochoska}, {Kohley}, {Kolenberg}, {Kontizas}, {Kontizas}, {Koposov},
{Kordopatis}, {Kostrzewa-Rutkowska}, {Koubsky}, {Lambert}, {Lanza}, {Lasne},
{Lavigne}, {Le Fustec}, {Le Poncin-Lafitte}, {Lebreton}, {Leccia}, {Leclerc},
{Lecoeur-Taibi}, {Lenhardt}, {Leroux}, {Liao}, {Licata}, {Lindstr{\o}m},
{Lister}, {Livanou}, {Lobel}, {L{\'o}pez}, {Managau}, {Mann}, {Mantelet},
{Marchal}, {Marchant}, {Marconi}, {Marinoni}, {Marschalk{\'o}}, {Marshall},
{Martino}, {Marton}, {Mary}, {Matijevi{\v{c}}}, {Mazeh}, {Messina},
{Michalik}, {Millar}, {Molina}, {Molinaro}, {Moln{\'a}r}, {Montegriffo},
{Mor}, {Morbidelli}, {Morel}, {Morris}, {Mulone}, {Muraveva}, {Musella},
{Nelemans}, {Nicastro}, {Noval}, {O'Mullane}, {Ord{\'e}novic},
{Ord{\'o}{\~n}ez-Blanco}, {Osborne}, {Pagani}, {Pagano}, {Pailler},
{Palacin}, {Palaversa}, {Panahi}, {Pawlak}, {Piersimoni}, {Pineau}, {Plachy},
{Plum}, {Poggio}, {Poujoulet}, {Pr{\v{s}}a}, {Pulone}, {Racero}, {Ragaini},
{Rambaux}, {Ramos-Lerate}, {Regibo}, {Riclet}, {Ripepi}, {Riva}, {Rivard},
{Rixon}, {Roegiers}, {Roelens}, {Romero-G{\'o}mez}, {Rowell}, {Royer},
{Ruiz-Dern}, {Sadowski}, {Sagrist{\`a} Sell{\'e}s}, {Sahlmann}, {Salgado},
{Salguero}, {Sanna}, {Santana-Ros}, {Sarasso}, {Savietto}, {Schultheis},
{Sciacca}, {Segol}, {Segovia}, {S{\'e}gransan}, {Shih}, {Siltala}, {Silva},
{Smart}, {Smith}, {Solano}, {Solitro}, {Sordo}, {Soria Nieto}, {Souchay},
{Spagna}, {Spoto}, {Stampa}, {Steele}, {Steidelm{\"u}ller}, {Stephenson},
{Stoev}, {Suess}, {Surdej}, {Szabados}, {Szegedi-Elek}, {Tapiador}, {Taris},
{Tauran}, {Taylor}, {Teixeira}, {Terrett}, {Teyssandier}, {Thuillot},
{Titarenko}, {Torra Clotet}, {Turon}, {Ulla}, {Utrilla}, {Uzzi}, {Vaillant},
{Valentini}, {Valette}, {van Elteren}, {Van Hemelryck}, {van Leeuwen},
{Vaschetto}, {Vecchiato}, {Viala}, {Vicente}, {Vogt}, {von Essen}, {Voss},
{Votruba}, {Voutsinas}, {Walmsley}, {Weiler}, {Wertz}, {Wevems},
{Wyrzykowski}, {Yoldas}, {{\v{Z}}erjal}, {Ziaeepour}, {Zorec}, {Zschocke},
{Zucker}, {Zurbach}, and {Zwitter}]{Gaia:2018}
{Gaia Collaboration}.; {Helmi}, A.; {van Leeuwen}, F.; {McMillan}, P.J.;
{Massari}, D.; {Antoja}, T.; {Robin}, A.C.; {Lindegren}, L.; {Bastian}, U.;
{Arenou}, F.;  et~al.
\newblock {Gaia Data Release 2. Kinematics of globular clusters and dwarf
galaxies around the Milky Way}.
\newblock {\em Astron. Astrophys.} {\bf 2018}, {\em 616},~A12.
 [\href{http://dx.doi.org/10.1051/0004-6361/201832698}{CrossRef}]

\bibitem[{Donoso} {et~al.}(2006){Donoso}, {O'Mill}, and
{Lambas}]{Donoso:2006}
{Donoso}, E.; {O'Mill}, A.; {Lambas}, D.G.
\newblock {Alignment between luminous red galaxies and surrounding structures
at $z\sim$0.5}.
\newblock {\em Mon. Not. R. Astron. Soc.} {\bf 2006}, {\em 369},~479--484.
 [\href{http://dx.doi.org/10.1111/j.1365-2966.2006.10328.x}{CrossRef}]

\bibitem[{Wang} {et~al.}(2010){Wang}, {Park}, {Hwang}, and
{Chen}]{Wang:2010}
{Wang}, Y.; {Park}, C.; {Hwang}, H.S.; {Chen}, X.
\newblock {Distribution of Satellite Galaxies in High-redshift Groups}.
\newblock {\em Astrophys. J.} {\bf 2010}, {\em 718},~762--767.
 [\href{http://dx.doi.org/10.1088/0004-637X/718/2/762}{CrossRef}]

\bibitem[{Gonzalez} {et~al.}(2023){Gonzalez}, {Zhao}, {Tang}, and
{Mathews}]{Gonzalez23}
{Gonzalez}, P.; {Zhao}, X.; {Tang}, G.; {Mathews}, G.
\newblock {Are Satellite Planes Around Milky Way-Like Galaxies Just Showing Us
Accretion Along The Filaments?}
\newblock In \emph{American Astronomical Society Meeting
Abstracts}; {American Astronomical Society Publishing: Washington, DC, USA}, 2023;  Volume~55, p. 354.01.

\bibitem[{Cooper} {et~al.}(2023){Cooper}, {Koposov}, {Allende Prieto},
{Manser}, {Kizhuprakkat}, {Myers}, {Dey}, {G{\"a}nsicke}, {Li}, {Rockosi},
{Valluri}, {Najita}, {Deason}, {Raichoor}, {Wang}, {Ting}, {Kim}, {Carrillo},
{Wang}, {Beraldo e Silva}, {Han}, {Ding}, {S{\'a}nchez-Conde}, {Aguilar},
{Ahlen}, {Bailey}, {Belokurov}, {Brooks}, {Cunha}, {Dawson}, {de la Macorra},
{Doel}, {Eisenstein}, {Fagrelius}, {Fanning}, {Font-Ribera}, {Forero-Romero},
{Gazta{\~n}aga}, {Gontcho a Gontcho}, {Guy}, {Honscheid}, {Kehoe}, {Kisner},
{Kremin}, {Landriau}, {Levi}, {Martini}, {Meisner}, {Miquel}, {Moustakas},
{Nie}, {Palanque-Delabrouille}, {Percival}, {Poppett}, {Prada}, {Rehemtulla},
{Schlafly}, {Schlegel}, {Schubnell}, {Sharples}, {Tarl{\'e}}, {Wechsler},
{Weinberg}, {Zhou}, and {Zou}]{Cooper:2023}
{Cooper}, A.P.; {Koposov}, S.E.; {Allende Prieto}, C.; {Manser}, C.J.;
{Kizhuprakkat}, N.; {Myers}, A.D.; {Dey}, A.; {G{\"a}nsicke}, B.T.; {Li},
T.S.; {Rockosi}, C.;  et~al.
\newblock {Overview of the DESI Milky Way Survey}.
\newblock {\em Astrophys. J.} {\bf 2023}, {\em 947},~37.
 [\href{http://dx.doi.org/10.3847/1538-4357/acb3c0}{CrossRef}]

\end{thebibliography}
\end{document}